\documentclass[journal]{IEEEtran}

\usepackage{hyperref}
\usepackage{amsmath,amssymb,amsfonts}
\usepackage{graphicx}
\usepackage{textcomp}
\usepackage[table]{xcolor}
\usepackage{subcaption}
\usepackage{gensymb}
\usepackage{float}
\usepackage{algorithm,algcompatible,algpseudocode}
\usepackage{array}
\usepackage{eso-pic, rotating, graphicx}
\usepackage{multirow}
\usepackage{multicol}
\usepackage{tabularx, makecell}
\usepackage[utf8]{inputenc}
\usepackage{svg}
\usepackage{ctable}
\usepackage{booktabs}
\usepackage{lineno}

\begin{document}

\title{Self-supervised Learning for Clustering of Wireless Spectrum Activity\\
}

\author{
\IEEEauthorblockN{Ljupcho Milosheski\IEEEauthorrefmark{1}\IEEEauthorrefmark{2},
Gregor Cerar\IEEEauthorrefmark{1},
Bla\v{z} Bertalani\v{c}\IEEEauthorrefmark{1},
Carolina Fortuna\IEEEauthorrefmark{1},
Mihael Mohorčič\IEEEauthorrefmark{1}\IEEEauthorrefmark{2}}\\
\IEEEauthorblockA{
\IEEEauthorrefmark{1}Department of Communication Technologies, Jozef Stefan Institute, Jamova 39, 1000, Ljubljana, Slovenia \\
\IEEEauthorrefmark{2}Jozef Stefan International Postgraduate School, Jamova 39, 1000, Ljubljana, Slovenia \\
Email: \{ljupcho.milosheski, miha.mohorcic, carolina.fortuna\}@ijs.si
}
}

\maketitle

\begin{abstract}
In recent years, much work has been done on processing of wireless spectrum data involving machine learning techniques in domain-related problems for cognitive radio networks, such as anomaly detection, modulation classification, technology classification and device fingerprinting. Most of the solutions are based on labeled data, created in a controlled manner and processed with supervised learning approaches.
However, spectrum data measured in real-world environment is highly nondeterministic, making its labeling a laborious and expensive process, requiring domain expertise, thus being one of the main drawbacks of using supervised learning approaches in this domain. 
In this paper, we investigate the {utilization} of self-supervised learning (SSL) for exploring spectrum activities in a real-world unlabeled data. In particular, we {assess} the performance of SSL models, based on the reference DeepCluster architecture. {We carefully consider the current state-of-the-art feature extractors, taking into account the performance and complexity trade-offs. Our findings demonstrate that} SSL models achieve superior performance regarding the {feature quality} and clustering performance {compared to baseline feature learning approaches}. With SSL models we achieve {significant} reduction of the feature vectors size by two orders of magnitude, while improving the performance by a factor {ranging from} 2 to 2.5 across the evaluation metrics, supported by visual assessment. {Furthermore, we showcase how adapting the reference SSL architecture to domain-specific data is followed by a substantial reduction in model complexity up to one order of magnitude, without compromising, and in some cases, even improving the clustering performance.}

\end{abstract}



\section{Introduction}

The number and type of wireless devices connected to the Internet is rapidly increasing with the current affordable personal mobile and Internet of Things (IoT) devices, requiring wireless networks to handle large number of connections and high traffic loads. As a reference, the requirement for the number of connected devices in the fifth-generation (5G) networks is one million devices per square kilometer. The existence of such a number of devices requires complex wireless resource management. Over time, several new approaches to wireless resource sharing, including dynamic spectrum access~\cite{zhao2007survey}, licensed shared access~\cite{frascolla2016dynamic} have been proposed. However, additional technological components, such as spectrum usage databases~\cite{chen2013database} and radio environment maps~\cite{kulzer2021cdi}, had to be developed to enable such sophisticated and dynamic spectrum usage approaches. To be able to correctly inform on spectrum usage, additional knowledge of other devices operating within the range of a wireless device is critical for future smart usage of the spectrum. In this respect, some of the recent efforts were focused on detecting the  modulation used~\cite{rajendran2018deep}, technology used~\cite{fontaine2019towards}, anomalous activities~\cite{tandiya2018deep}, etc.  

As also discussed in~\cite{zhang2019deep} and~\cite{wong2020rfml}, significant effort is still being invested in the field to develop accurate and scalable deep learning (DL) algorithms able to accurately and automatically manage spectrum resource usage. With respect to the learning approach, these techniques can be divided into (1)~\textit{supervised} that require labels to be present for the training data, and (2)~\textit{unsupervised} that do not assume any such labels. Applications in wireless spectrum management need to be aware of operating details (i.e., type of technology, transmission parameters, etc.). Development of a DL-based model to support such application typically requires labelled data that is expensive to acquire as it requires complex wireless and computing equipment~\cite{rajendran2017electrosense, vsolc2015low} or intense labelling efforts by domain experts~\cite{gale2020automatic} that do not always lead to high quality labels due to the nondeterministic nature of wireless operating environments. \textit{Semi-supervised} and \textit{active-learning} emerged as alternative techniques that have the advantage of using a relatively small amount of labeled samples for achieving performance that is comparable to the regular supervised approach. 

Given the advent of large datasets which are expensive or practically impossible to label, \textit{self-supervised learning} (SSL)~\cite{jing2020self}, as another intermediate learning approach, is becoming an important alternative that is particularly suitable to reduce the data labelling cost and leverage the unlabelled data pool. SSL is a representation learning method where a supervised task is created out of the unlabelled data. Using an SSL approach, it is possible to create very similar groups (i.e. clusters) from a large, unlabelled dataset and then label each cluster. By labelling the learnt clusters, it is possible to then use the model as a classifier by assigning new, unseen examples to those clusters and therefore label them as one would do in a typical classification task.

Developing an easy to use, automated and technology agnostic way to explore spectrum activities and group similar activities, eventually enabling automatic rather than manual transmission identification and cataloguing as currently done for instance in the Signal Identification Guide\footnote{\url{https://www.sigidwiki.com/wiki/Signal_Identification_Guide}}, is still an open research topic which motivated this investigation.

{Employing DL architectures based on convolutional neural networks (CNN) allows for direct processing and classification of the raw In-phase and Quadrature (I/Q) time series \cite{o2018over}. While for modulation classification \cite{rajendran2018deep} and device fingerprinting \cite{robinson2020dilated}, direct processing on the raw I/Q time series data is desirable, for radio access technology classification it is better to employ spectrograms due to lower complexity and robustness towards low SNR conditions \cite{fontaine2019towards}. Additionally, the image-like (2D matrices) format of the spectrograms provide the possibility to utilize some state-of-the-art architectures from the closely related machine vision domain, such as Vision Transformers (ViT) \cite{dosovitskiy2020image} that recently emerged as competitors to the CNN-based solutions. Inter domain adaptations, such as in \cite{fonseca2021radio}, could additionally stimulate the development of spectrum sensing technologies which are increasingly important for the next generation radio networks.}

In this paper, we investigate the suitability of SSL to support automatic spectrum exploration on an example of an unlicensed 868 MHz Short Range Device (SRD) band in an urban environment using 15 days of spectrum sweeps collected in {the} LOG-a-TEC\footnote{\url{https://log-a-tec.eu/datasets.html}} {wireless testbed}. Leveraging this data, we propose an SSL architecture adapted for spectrum activity identification and clustering, in which segments of spectrograms containing signal activity are used to train the DL self-supervised network and enable the discovery of the types of transmissions available over the respective period of time. It is based on machine vision and inspired by DeepCluster~\cite{caron2018deep}, which is also used in this study as a reference SSL model. {In this architecture, we experiment with CNN-based and ViT-based feature extraction and} prove that such an architecture is suitable for spectrogram analysis by learning spectrum features and clustering spectrograms based on their content.

The main contributions of this work can be summarised as follows:
\begin{itemize}
    \item Adaptation of an SSL architecture to discover wireless transmissions in real-world spectrogram data when no prior knowledge (i.e. labels) is available, while also achieving significant reduction of the complexity of the architecture with {up to 10} times less trainable parameters compared to the {original implementation}.
    \item Proposing dimensionality reduction of the {embedding space by principal component analysis (PCA)} using a threshold on the amount of explained variance ratio (EVR), achieving features quality and clustering performance improvement by a factor of 2-2.5 across the selected evaluation metrics.
    {\item Development of a methodology for quantitative and qualitative evaluation of the transmissions clustering from raw spectrograms. The methodology consists of features quality assessment in the embedded space and clustering evaluation with selected metrics, verified by data-specific visualizations.}
    \item Experimental evaluation of the adapted SSL architecture for two use cases, activity detection and fine grain transmissions classification.
    {\item Comparison of two feature extraction modules in the adapted SSL architecture, one based on ViT, considered as more powerful in RGB image classification tasks, and CNN, that proved more suitable for low-content data such as spectrograms, taking into account the performance and complexity trade-offs.}
\end{itemize}

The rest of the paper is structured as follows. Section \ref{sec:related} analyzes the related work, Section \ref{sec:Architecture} introduces SSL and baseline system architectures, Section \ref{sec:methodology} elaborates on the experimental methodology and Section \ref{sec:Results} presents the experimental results. Finally, Section \ref{sec:concl} concludes the paper.

\begin{table*}[hbt!]
\centering
\scriptsize
\caption{Related works with main characteristics}
\label{tab:Related}
\begin{tabular}{p{0.05\textwidth}p{0.15\textwidth}p{0.07\textwidth}p{0.07\textwidth}p{0.1\textwidth}p{0.1\textwidth}p{0.05\textwidth}p{0.1\textwidth}p{0.1\textwidth}}
    \toprule
    \textbf{Publi-cation} 
    & \textbf{Problem type} 
    & \textbf{Architec-ture}
    & \textbf{Data type}
    & \textbf{Dataset}
    & \textbf{Conti-nuous sensing (yes/no)}
    & \textbf{Labels (yes/no)}
    & \textbf{Approach}
    & \textbf{Band} \\
    \midrule
    \cite{{rajendran2018deep}}
    & Modulation classification \textit{(multiclass)} 
    & LSTM
    & I/Q, PSD
    & RadioML, Electrosense
    & Yes
    & Yes
    & Supervised
    & 174-230 MHz, 470-862MHz, 25-1300 MHz \\
    \midrule
    \cite{kuzdeba2021transfer}
    & Device fingerprinting \textit{(multiclass)}
    & DCC
    & I/Q
    & DARPA, Synthetic
    & /
    & Yes
    & Supervised
    & 2.4GHz, 5GHz, 978MHz, 1090MHz \\
    \midrule
    \cite{riyaz2018deep}
    & Device fingerprinting \textit{(multiclass)}
    & CNN
    & I/Q
    & Test-bed
    & /
    & Yes
    & Supervised
    & 5GHz \\
    \midrule
    \cite{fontaine2019towards}
    & Technology classification \textit{(multiclass)}
    & DT, FNN, CNN
    & RSSI, I/Q, Spectro-grams
    & Ghent, Dublin
    & Yes
    & Yes
    & Supervised
    & 5540MHz, 2412MHz, 806MHz \\
    \midrule
    \cite{o2018over}
    & Modulation classification \textit{(multiclass)}
    & CNN
    & I/Q
    & Test-bed, synthetic
    & /
    & Yes
    & Supervised
    & 900MHz \\
    \midrule
    \cite{fonseca2021radio}
    & Technology characterization trough object detection \textit{(multiclass)}
    & YOLO CNN
    & Spectro-grams
    & Test-bed, Ghent
    & Yes
    & Yes
    & Supervised
    & 5540MHz, 2412MHz, 806MHz \\
    \midrule
    \cite{wong2018clustering}
    & Device fingerprinting with clustering \textit{(multiclass)}
    & CNN
    & I/Q
    & Test-bed
    & /
    & Yes
    & Semi-supervised
    & 0.25 - 1.25MHz, 1.67MHz, 2.5MHz \\
    \midrule
    \cite{robinson2020dilated} 
    & Device fingerprinting \textit{(multiclass)}
    & DCC + CNN
    & I/Q
    & DARPA
    & /
    & Yes
    & Supervised, Unsupervised
    & 2.4GHz, 5GHz, 978MHz, 1090MHz \\
    \midrule
    \cite{tandiya2018deep}
    & Anomaly detection \textit{(binary)}
    & PredNet-autoencoder
    & Spectro-grams
    & Synthetic
    & /
    & No
    & Unsupervised*
    & / \\
    \midrule
    \cite{rajendran2018saife}
    & Anomaly detection \textit{(binary)}
    & Supervised CNN-based
    & PSD
    & Synthetic, HackRF SDR, Electrosense
    & Yes
    & Yes
    & Unsupervised
    & 10MHz-3GHz \\
    \midrule
    \cite{gale2020automatic}
    & Transmissions detection and classification \textit{(binary)}
    & Image processing
    & Spectro-grams
    & Log-A-Tec
    & Yes
    & Yes
    & Unsupervised
    & 868MHz \\
    \bottomrule
    Ours
    & Transmissions clustering \textit{(multiclass)}
    & Deep-clustering
    & Spectro-grams
    & Log-A-Tec
    & Yes
    & No
    & Unsupervised (Self-supervised)
    & 868MHz \\
    \bottomrule
\end{tabular}	
\end{table*}

\section{Related work}
\label{sec:related}



In recent years, as in many other research areas the use of DL models gained a lot of attention also in the development of algorithms for processing spectrum data. Selected works from the domain which are considered as most relevant and closely related to this paper are listed in Table~\ref{tab:Related}, which summarizes their main characteristics, and briefly elaborate in the following where they are grouped according to the adopted learning approach. 

\subsection{Supervised learning}

In the existing works, \textit{supervised} DL-based models are most widely represented and they achieve significant increase of performance when compared to more traditional ML approaches, as shown in \cite{fontaine2019towards}, \cite{o2018over}, \cite{riyaz2018deep}.

In~\cite{fontaine2019towards}, the authors compare the performance and generalization ability of models that use manually extracted expert features with models that use raw spectrum data. They are solving the task of technology classification on a dataset containing transmissions of three different wireless technologies. They prove that using CNN on raw I/Q data or spectrogram images outperforms all other models in terms of accuracy, generalization ability to unseen datasets from different operating environments, and robustness to different noise levels.

In~\cite{riyaz2018deep}, an application of CNN supervised learning for device identification, again using raw I/Q data, is proposed. The dataset contains transmissions from five devices. Device identification is based on the CNN's ability to learn various device-specific impairments in the raw signals. SVM and logistic regression are used as reference models for performance comparison and authors show that the proposed CNN model significantly outperforms the baseline algorithms for the posed task. 

In many cases, state-of-the-art accuracy is achieved by adopting and modifying DL-based architectures that are already known and well established in other, closely related signal processing fields, such as image and sound processing \cite{kuzdeba2021transfer}, \cite{tandiya2018deep}. In \cite{robinson2020dilated}, device classification task using real-world I/Q data of transmissions from a large population of nearly $10,000$ devices is solved using a new neural network architecture based on dilated causal convolutional layers. The design is motivated by an existing audio signals processing architecture.

\subsection{Semi-supervised and transfer learning}

Although the supervised models have superior performance judging by their accuracy, the necessity for large labeled training datasets as one of the major downsides of this approach remains. One way of going around this problem is to train feature extractor in unsupervised manner and then tune the classification on a small labelled dataset. This approach is employed in \cite{rajendran2018saife} to solve the problem of anomaly detection in wireless communications.

Another approach for the labeling problem is employing transfer learning \cite{kuzdeba2021transfer}, \cite{fonseca2021radio}. In \cite{fonseca2021radio}, object detection "You Only Look Once" (YOLO) model, pretrained on the ImageNet dataset, is tuned to detect and classify different types of transmissions using spectrograms. It is shown that the proposed model performs well in classifying interfering signals on simulated data with different signal-to-noise ratios and provides additional information about the position of the transmission events in the spectrum.

Although the semi-supervised and transfer learning are proven to lower the amount of data that is required for the training, labeled data is still needed for the tuning. Spectrum data content is not as universal as the RGB images and it is much more dependent on the operating environment regarding the noise, signal strength, fading, multipath effects, etc., which makes it necessary to provide labels for each specific radio environment.


\subsection{Unsupervised learning}

Considering the availability of large amounts of unlabelled spectrum data \cite{vsolc2015low}, \cite{rajendran2017electrosense}, another approach to the labeling problem with high potential, which we consider as under-explored in the spectrum data processing domain, is using unsupervised models. We believe that the development of unsupervised approaches could more effectively solve the labeling problems and provide highly automated architectures that can be adapted to different operating environments with minimum expert intervention. 

There are existing efforts in this direction, but they only consider the marginal cases of event detection (binary classification). A pipeline for automatic detection of wireless transmissions using classic image processing techniques is proposed in \cite{gale2020automatic}. Another such example is the work in \cite{feng2017anomaly} where auto-encoder is utilized as feature learner for anomaly detection and shown to outperform the robust PCA.

{Some recent works employ contrastive learning as an SSL representation learning approach, for instance on I/Q samples \cite{davaslioglu2022self} and motion sensor data \cite{liu2021contrastive}. Although they show promising performance, this approach is more suitable for datasets that have clear distinction between positive and negative samples, while also requiring data-specific transformations for training.}

In our work, we align with the efforts of developing  completely unsupervised models that would provide feature learning in an automated manner. We develop a self-supervised architecture that can provide the functionality of simple activity detection, but also enable fine grain classification of wireless transmissions of different technologies in wireless spectrum data.

\section{Architectures of the SSL and baseline systems} 
\label{sec:Architecture}

{In this paper we investigate the suitability of SSL deep clustering models for automatic learning of relevant features and clustering wireless spectrum activities identified from spectrogram segments. These activities represent various transmission events, each characterized by distinct shapes and signal intensities. The learning process relies on an iterative refinement of the feature extraction parameters, in our case in modules based on CNN or ViT, and clustering assignments which are updated in alternating manner. No prior knowledge is assumed about the types and number of different spectrum activities present in the dataset or their specific visual appearance in the spectrograms. Since the spectrograms are 2D data, similar to gray-scale images, we follow the same ideas as in \cite{caron2018deep}. }
{We conduct a comparative analysis of Self-Supervised Learning (SSL) models against two distinct baseline approaches for feature learning. The first baseline method employs PCA as a fundamental unsupervised technique for representation learning. The second baseline utilizes a CNN-based auto-encoder, which represents a more advanced unsupervised DL approach in our evaluation.}

\begin{figure*}[hbt!]
\includegraphics[width=\textwidth]{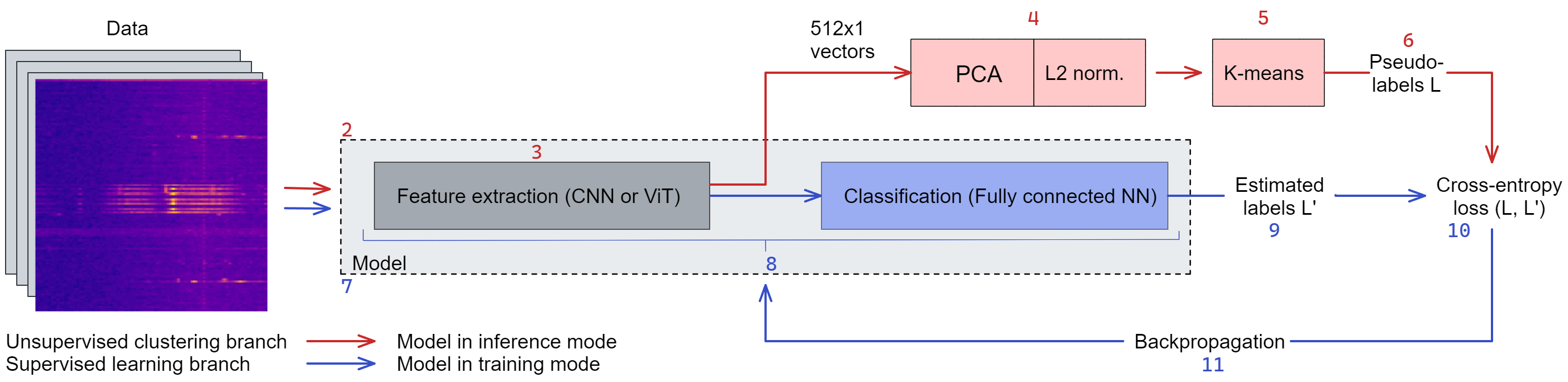}
\caption[width=\textwidth]{Architecture of the self-supervised CNN system.}
\label{fig:ArchitectureDiagram}
\end{figure*}

\subsection{SSL architectures}
\label{sec:ssmodel}

As a reference SSL system we adopted the existing self-supervised DeepCluster architecture proposed in \cite{caron2018deep} for RGB image feature learning based on a VGG (Visual Geometry Group) standard deep Convolutional Neural Network (CNN) architecture with batch normalization that is regularly used in computer vision applications. This system alternates between clustering the image features produced by CNN and updating its weights by predicting the cluster assignments.

Inspired by the approach of this reference system, we developed a SSL system depicted in Figure~\ref{fig:ArchitectureDiagram}, which is actually an adaptation of DeepCluster for spectrum activity identification and clustering from spectrograms, characterized by much less content compared to RGB images, thus allowing for significant complexity reduction and performance optimization. As it can be seen from the figure, the system design contains two branches, (1) a so-called unsupervised branch depicted with red arrows and (2) a supervised branch depicted with blue arrows. Both branches contain a feature extractor. In the unsupervised branch it is followed by a dimensionality reduction (PCA) and normalization (L2) preprocessing block and concluded with the K-means clustering block. In the supervised branch, the feature extractor is followed by fully connected layers as a classifier. Note that even if we refer to the second branch as supervised, it does not rely on actual labelled data. It relies on pseudo-labels that are repeatedly generated by K-means in unsupervised branch and refined through feedback, i.e. backpropagation. By working in tandem, the two branches realize self-supervision.

Compared to the original DeepCluster architecture, we propose the following variations:

\paragraph*{Alternative CNN}
{We want to better understand the discriminative power of CNNs which should provide feature vectors in the embedding space that are correlated to the transmission shapes, and utilize clustering algorithm in the embedding space that will provide pseudo labels for CNN's parameters optimization, together resulting in iterative learning and grouping of the samples. }

Motivated by the findings in \cite{o2018over}, we also select ResNet architecture as a possible feature extraction architecture that can replace VGG in the DeepCluster architecture depicted in Figure \ref{fig:ArchitectureDiagram} as it was shown to perform better in a supervised task of modulation classification. As we show later, with ResNet we achieve very similar performance as with VGG at significant complexity reduction of roughly 10 times less trainable parameters. From the ResNet family we use ResNet18 (18 showing the number of convolution layers in CNN) in its original form, but with input and output layers customized according to the shape of the images and the number of classes, i.e., a single input channel in the case of spectrogram images compared to a 3-channel input required for RGB images that ResNet18 was originally designed for.

We also added PCA feature space reduction before clustering to adapt to the significantly lower amount of content in spectrograms compared to RGB images.

\paragraph*{Transformers}
{A more recent and very powerful architecture in natural language processing and machine vision is represented by transformers, therefore we also we also consider ViT in the SSL architecture, as it is proven \cite{caron2021emerging} to be powerful feature extraction module in the machine vision domain.}

{ViT-based models are devised by simple replacement of the feature extraction and classification modules of the SSL architecture with transformer architecture, which includes projection head (classification module) itself. In such model, features from the ViT embedded space undergo the same process of PCA-transformation and L2 normalization before being clustered by the K-means.}

{Further in the paper, we used the following template notation AA-FE-X where AA refers to the architectural approach, and can have two values: SSL or B from baseline. FE stands for feature extractor and can have values VGG, RN from ResNet, ViT, AE, PCA while X represents the number of PCA components used for the model instance development. For example, we refer to the VGG-based and ResNet-based self-supervised models with 20 PCA components as SSL-VGG-20 and SSL-RN-20, respectively.}

The pseudo-code corresponding to the workflow of the proposed self-supervised system is  given in Algorithm \ref{alg:Algorithm}. For clarity and clear mapping, the blocks of the system in Figure \ref{fig:ArchitectureDiagram} are marked with the corresponding line  numbers from the pseudo-code in Algorithm \ref{alg:Algorithm}. 

\subsubsection{Workflow of the unsupervised branch}
In the initial phase, the CNN{/ViT} is initialized randomly and the input data, which consists of image-like spectrum segments, is unlabeled. The clustering algorithm (unsupervised branch, marked with red arrows in Figure \ref{fig:ArchitectureDiagram}) is used to cluster the features and provide pseudo-labels (denoted by L in the figure) at the beginning of each training epoch. Initialization of the cluster centers is random. The features are extracted using the {feature extraction module}, which has the size of $512 \times 1$ {for the CNN and $1024 \times 1$ for the ViT} (lines 2-3 in Algorithm \ref{alg:Algorithm}). Thus, a descriptor in the form of $512 \times 1$ or {$1024 \times 1$} vector is obtained for each spectrum segment image. These descriptors are then PCA-reduced, L2-normalized and finally clustered (lines 4-5 in Algorithm \ref{alg:Algorithm}).

\begin{algorithm}
    \caption{Workflow of the architecture}
    \begin{algorithmic}[1]
    \While{$Iteration$\textless $Number Of Epochs$}
        
        \COMMENT{\textbf{Unsupervised branch:}}
        \State $Model \gets Evaluation Mode$
        \State $Extract Features$ \Comment{CNN or ViT}
        \State $Process Features$ \Comment{PCA + L2 norm.}
        \State $Cluster Features$ \Comment{K-Means}
        \State $(L) Pseudo-labels \gets Cluster Assignments$  
        
        \COMMENT{\textbf{Supervised branch:}}
        \State $Model \gets Training Mode$
        \State $Estimate Labels$
        \State $L' \gets Estimated Labels$
        \State $Cross-entropy(L, L')$
        \State $Backpropagation$ \Comment{Update classifier weights}
    \EndWhile{}
    \end{algorithmic}
    \label{alg:Algorithm}
\end{algorithm}

\subsubsection{Working of the supervised branch}
In this phase, {the feature extraction and classification modules are} used in training mode, as a supervised classification model. The flow of this pipeline activity is indicated by the blue arrows in Figure \ref{fig:ArchitectureDiagram}. Using the provided cluster assignments from the previous step as pseudo-labels (L) for the input images, {model} is trained for one epoch. This completes one iteration of the entire pipeline work cycle. 

The procedure stops when the predefined number of iterations (training epochs) is reached. In our experiments, we used 200 training epochs. This number was determined empirically by observing the convergence of the loss function.

\begin{figure}[hbt!]
    \begin{subfigure}{\columnwidth}
        \centering
        \includegraphics[width=\columnwidth]{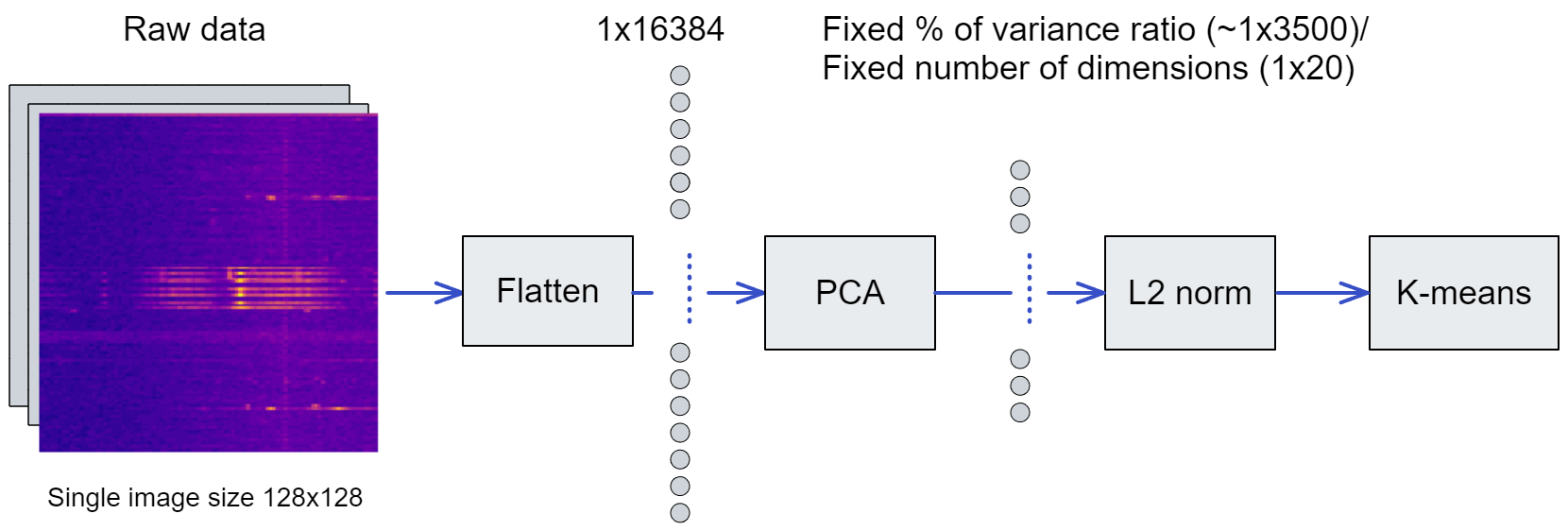}
        \caption{PCA baseline system.}
        \label{fig:BaseAlgProc}
    \end{subfigure}
    \begin{subfigure}{\columnwidth}
        \centering
        \includegraphics[width=\columnwidth]{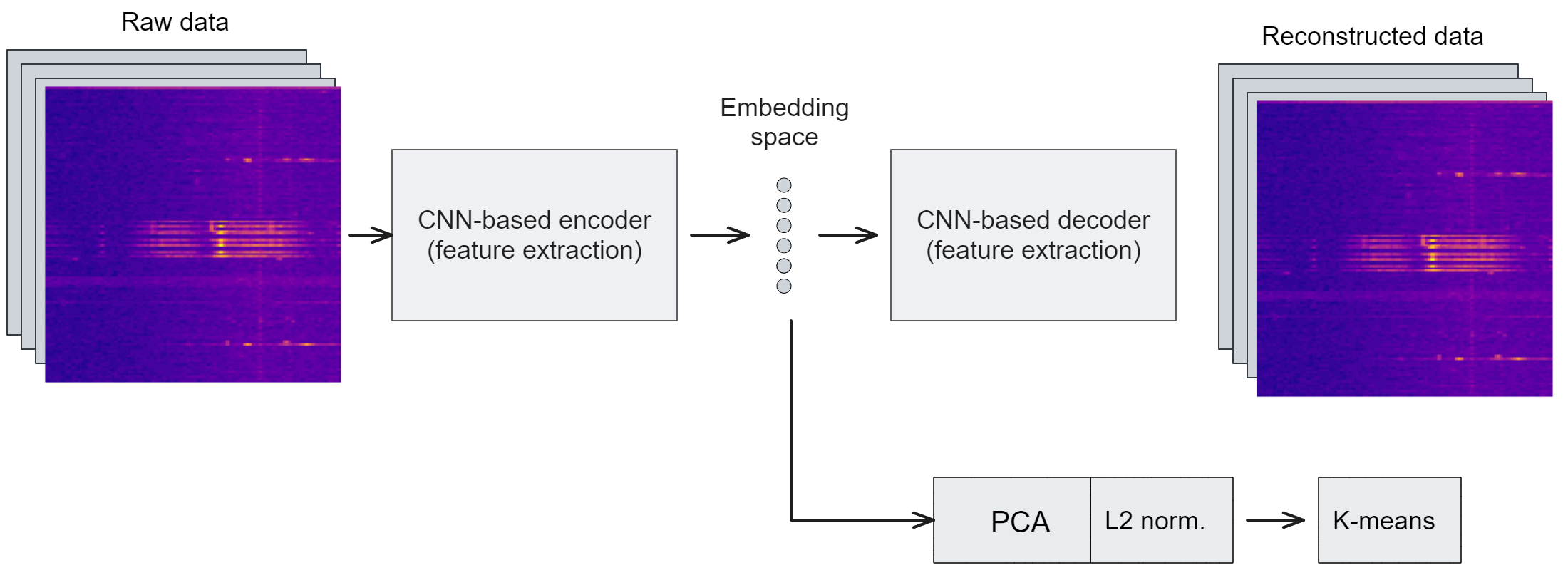}
        \caption{{Auto-encoder baseline system.}}
        \label{fig:BaseAE}
    \end{subfigure}
    \caption{{Architectures of the baseline systems.}}
    \label{fig:BaselineSystems}
\end{figure}

\subsection{Baseline systems}
\label{sec:bmodel}

Considering that the most influential part of the architecture is the feature learning component, we take PCA-based representation learning as {a simple, yet informative} baseline for comparison with the SSL models complemented by a K-means clustering algorithm, as depicted in Figure \ref{fig:BaseAlgProc}.
The system consists of a flattening block, PCA-dimensionality reduction and L2 normalization for input data preprocessing, followed by the K-means clustering block.

The flattening reorders the elements of the input matrix into a single row for each data sample. On the vectors provided by the flattening, the same sequence of operations is applied, as for the vectors provided by the self-supervised system. PCA is applied on the flattened data and as a result of this operation, reduced feature vectors representing the input data are obtained. These vectors are then L2-normalized and finally used as an input to the K-means clustering algorithm. In the rest of the paper we refer to baseline models as B-PCA.

{As a second, more comprehensive baseline algorithm for comparison we utilize a symmetric CNN-based auto-encoder (AE) architecture realized by the same ResNet18 module, which was proposed as feature extractor in the SSL architecture. The visualization of the system is depicted in Figure \ref{fig:BaseAE}. In the same manner as for the SSL models, PCA is being applied on the embedding space feature vectors to provide reduced dimensionality space, which is later L2 normalized and than fed to the clustering algorithm. Thus we provide stronger baseline model for comparison with similar feature extraction module as the proposed SSL. The training, same as for the SSL model, was performed for $200$ epochs.}

\section{Methodology}
\label{sec:methodology}
In view of developing and benchmarking unsupervised spectrum activity exploration we define the methodological approach to prepare the raw data used to train the proposed self-supervised and baseline models, and to develop and evaluate high quality feature extraction, clustering and resulting models. 

\subsection{Raw training data preparation}\label{sec:Data}
The dataset used for the analysis consists of fifteen days of spectrum measurements acquired at a sampling rate of 5 power spectrum density (PSD) measurements per second using 1024 FFT bins in the 868\,MHz license-free (shared spectrum) SRD band with a 192\,kHz bandwidth. The data is acquired in the LOG-a-TEC testbed. Details of the acquisition process and a subset of the data can be found in \cite{vsolc2015low}. The acquired data has a matrix form of $1024 \times N$, where N is the number of measurements over time. By windowing the data with a window size W, the resulting raw images for training would have $1024 \times W$ dimension which can be computationally untractable for less capable computing platforms. Therefore, in addition to windowing in time, we are also windowing the dataset in frequency (i.e., FFT bins) direction.

\begin{figure*}[t!]
\noindent\makebox[\textwidth]{\includegraphics[width=\textwidth]{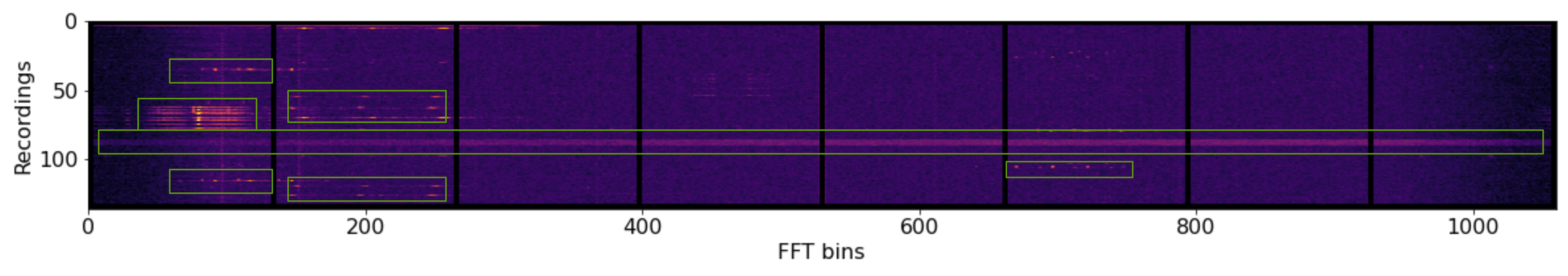}}
\caption[width=\textwidth]{Sample of 8 spectrogram segments from the data.}
\label{fig:SampleData}
\end{figure*}

The segmentation of the complete data-matrix into non-overlapping square images along time and frequency (FFT bins) is realized for a window size $W=128$. An example of such segmentation  containing 8 square images is shown in Figure \ref{fig:SampleData}, corresponding to image resolution of 25.6 seconds (128 measurements taken at 5 measurements per second) by 24 kHz. The window size is chosen to be large enough to contain any single type of activity and small enough to avoid having too many activities in a single image while also having in mind computational cost. Dividing the entire dataset of 15 days using $W=128$ and zero overlapping, produces 423,904 images of $128 \times 128$ pixels, where the pixel values are scaled to the range between 0 and 1.

From previous work \cite{gale2020automatic} we obtained labels of part of the transmissions in the employed dataset provided by experts. The labels are bounding boxes around transmissions in the same spectrogram data. We adapt these labels to our use case, considering the data is segmented in square-shaped regions. Each label as a rectangular region in the data matrix covers at least one square segment. All of the square segments that are overlapping with the labeled regions are marked as active, i.e., containing transmission(s). Thus we obtained labeled subset of square-shaped spectrograms that contain transmissions, as marked by experts, which have been used for the subsequent evaluation of the compared models.

The VGG-based model also requires adaptation on the dataset itself. In fact, using the described data segmentation in the SSL-VGG model, which was originally designed for input data size of $224 \times 224$ pixels, leads to converging the model to create only one cluster and zero valued feature vectors. To avoid this, we upscale the $128 \times 128$ spectrograms to the originally required input size for SSL-VGG. Another issue with SSL-VGG model is the number of training epochs. Running for too many (e.g., more than 40) epochs again leads to the behaviour explained above. The number of epochs that provides meaningful results with the SSL-VGG model was experimentally determined to be around $30$, which is notably less than the number of epochs required for the SSL-RN model. We assume that the reason for such behaviour is the low amount of content in the spectrograms compared to the content-rich RGB images that the SSL-VGG architecture was originally designed for.
{On the other side, the other two SSL models (SSL-RN and the SSL-ViT) along with the baseline AE model do not require additional data adaptation and consistently improve their performance during training.}

\subsection{Feature development and evaluation methodology}
\label{sec:features}
As shown in Figure~\ref{fig:ArchitectureDiagram},the segmented and normalized raw data is passed through a feature extractor (ResNet18) and PCA/L2 preprocessing, before being actually clustered. These blocks, together, learn to engineer features. The better the quality of the engineered features that train the K-means clustering algorithm, the better the resulting clustering model. Therefore, in the process of developing the proposed self-supervised model, the feature development process needs to be tuned and evaluated. 

While the feature extractor is automatically trained, the PCA performing dimensionality reduction for both the proposed self-supervised and the baseline system needs to be configured. We first analyze how many PCA components are needed to keep the most relevant information while discarding as much noise as possible. The number of dimensions $D$ used in the PCA-reduced dimensionality representation is determined by setting a threshold on the amount of Explained Variance Ratio (EVR). 
We assume that the important information is encoded in much smaller number of dimensions (i.e., between 10 and 20 out of 512) because, contrary to the RGB images from the reference application of SSL model \cite{caron2018deep}, the spectrograms contain much smaller amount of content. Visual analysis of the spectrograms shows that the transmission bursts surface is varying relative to the spectrogram size, but it is not bigger than around 10\%, i.e., 1600 pixels of the spectrogram relative to its size of $128\times 128=16384$ pixels. The EVR threshold is determined by exploring the plot of the cumulative sum of EVR of the PCA-transformed space dimensions.

Clusterability of developed features can be evaluated with different metrics such as the Hopkins score \cite{hopkins1954new} and visual assessment of clustering tendency (VAT) \cite{bezdek2002vat}. The Hopkins score is a metric that shows the probability that randomly sampled subset of the data comes from a uniform distribution. 
Resulting values are between 0 and 1, with 0 meaning the data is not uniformly distributed. However, having nonuniform distribution does not guarantee existence of clusters in the data. One such case is if the data has normal distribution, which will show low Hopkins score but will not contain any meaningful clusters. To prevent possible false conclusions derived based on this metric alone, we also use the evaluation with the VAT algorithm, which produces matrix visualisation of the dissimilarity of samples based on their pairwise euclidean distances. The value of each element of the visualization matrix is proportional to the pairwise dissimilarity between each of the samples to all of the other samples. Thus, the left diagonal of the matrix is with zero values representing dissimilarity of each sample to itself. The samples are ordered in such a way that groups that are located close in the feature space, according to the distance metric, appear as dark squares along the diagonal of the matrix. Implementation wise, an improved version of the VAT (iVAT) is used, which provides better visualization than the standard one.

In the evaluation process, the Hopkins score and the VAT plots are complementary and will be considered both at once, since VAT does not provide quantification, while the Hopkins score alone only gives an information about the uniformity of data distribution and not on its clusterability.

\subsection{Cluster development and evaluation methodology} \label{sec:EvalClus}
To develop the cluster model that best finds similar transmissions, the compared architectures have to be trained several times for different values of $k$ in K-means resulting in as many models. These models then need to be evaluated to see which one contains the cleanest clusters, i.e. provides clear data separation. As all the possible transmissions that may occur are not known in advance, we choose $k\in{[2,3,4,...30]}$. 

The clustering is then evaluated with two standard metrics, the Silhouette score \cite{rousseeuw1987silhouettes} and the Davies-Bouldin index  \cite{davies1979cluster}, and also manually by cluster analysis and explanation.

The Silhouette score is a value that measures the quality of the clustering by evaluating how similar is each sample to its assigned cluster.
The values it takes are in the range of $[-1, 1]$, where bigger values mean better clustering. Silhouette score of a cluster is the average of the scores of its elements, while Silhouette score of a clustering is the average of the scores of all of the clusters.

The Davies-Bouldin index is a clustering quality metric calculated as average of the similarity value of each cluster to its most similar (closest) cluster.
The range of values that this metric takes has only lower bound $0$ and smaller values mean better clustering.

As discussed in Section \ref{sec:Architecture}, the clustering algorithm is randomly initialized, so labels are not fixed to any specific type of content. Thus, besides the quantitative evaluation of the quality of the clustering results, manual evaluation is also performed in order to interpret the content that is specific for each cluster together with the frequency that is specific for the content. The motivation for the manual inspection is to verify that the clusters formed as output of the K-means are meaningful subsets of spectrograms containing highly correlated types of spectrum patterns (transmissions). Evaluation is made on plots of averaged spectrograms and histograms of frequency locations specific for the cluster's samples.

\subsection{Evaluation with labeled data} \label{sec:EvalLabels}
After manual evaluation we can assign specific type of transmission to each of the formed clusters. Also, it is justified to expect that there will be clusters with empty/no transmission segments since the dataset is created by uninterrupted measurements from certain time period and it is hardly possible that each spectrogram will contain transmission. This is verified by visualization of sample data. Knowing this, we evaluate the models by utilizing the created labeled data, as explained in Section~\ref{sec:Data}, and check how well the models are performing for the task of activity detection. The evaluation is made by running the labeled spectrograms containing transmission trough the trained models and calculate the percentage of them that were correctly distributed across the clusters that contain transmissions, marked in the previous step of manual evaluation. Correct classifications would be labeled samples assigned to any of the transmisions-rich clusters and missclassifications would be labeled samples assigned to the empty/no activity clusters.

\begin{figure}[hbt!]
    \begin{subfigure}{\columnwidth}
        \centering
        \includegraphics[width=\columnwidth]{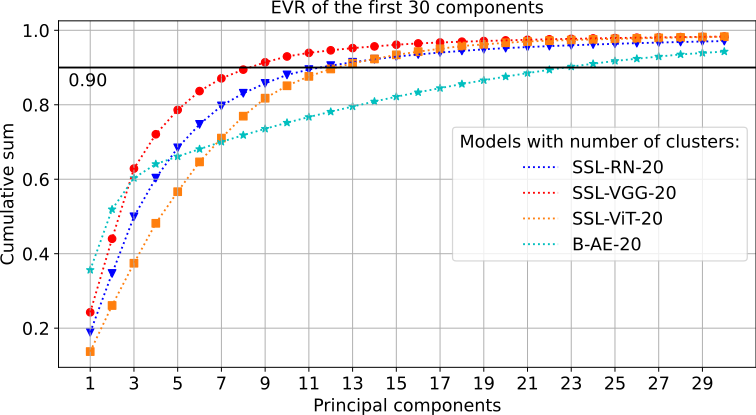}
        \caption{PCA on CNN extracted features.}
        \label{fig:PCAonCNN}
    \end{subfigure}
    \begin{subfigure}{\columnwidth}
        \centering
        \includegraphics[width=\columnwidth]{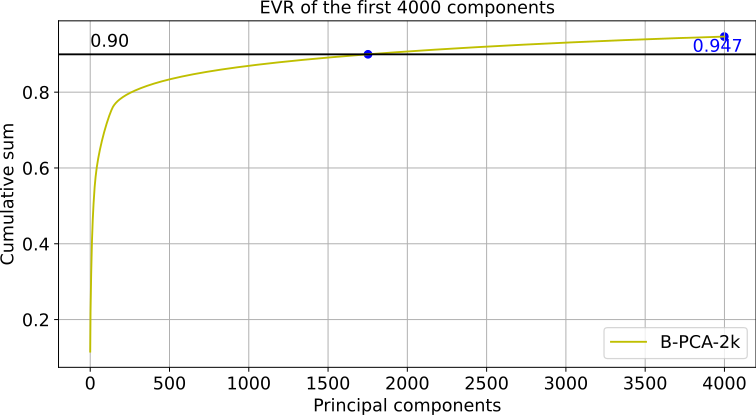}
        \caption{Direct PCA on images for the baseline model.}
        \label{fig:DirectPCAonImages}
    \end{subfigure}
    \caption{{Cumulative sum of EVR of the PCA-transformed features.}}
    \label{fig:VarRatio}
\end{figure}

\begin{figure*}[hbt!]
    \centering
    \begin{subfigure}[b]{0.16\textwidth}
         \centering
         \includegraphics[width=\textwidth]{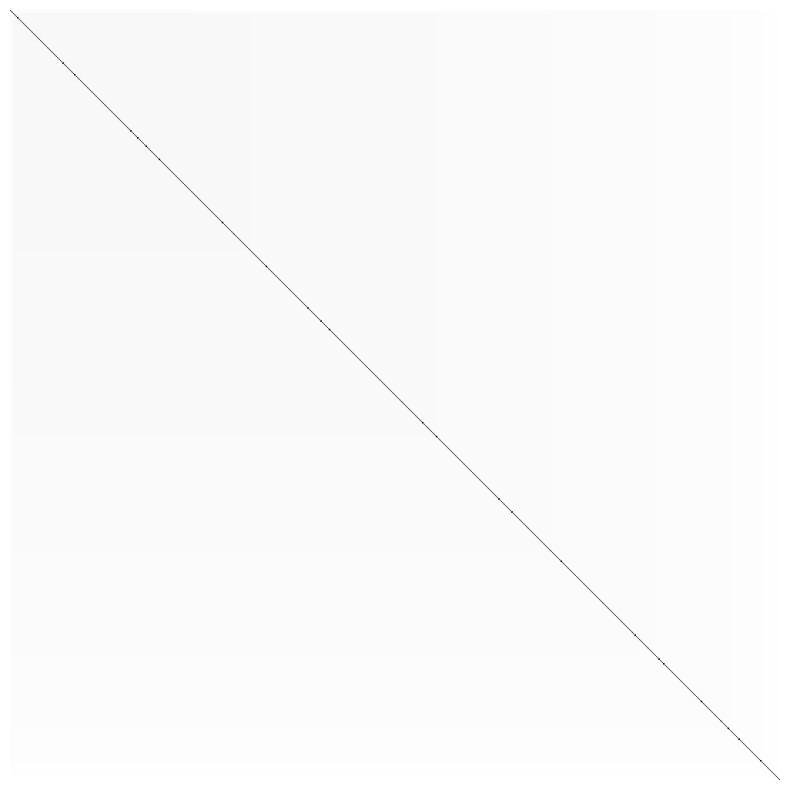}
         \caption{B-PCA-2k}
         \label{fig:VATVarRatio95}
    \end{subfigure}
    \hfill
    \begin{subfigure}[b]{0.16\textwidth}
        \centering
        \includegraphics[width=\textwidth]{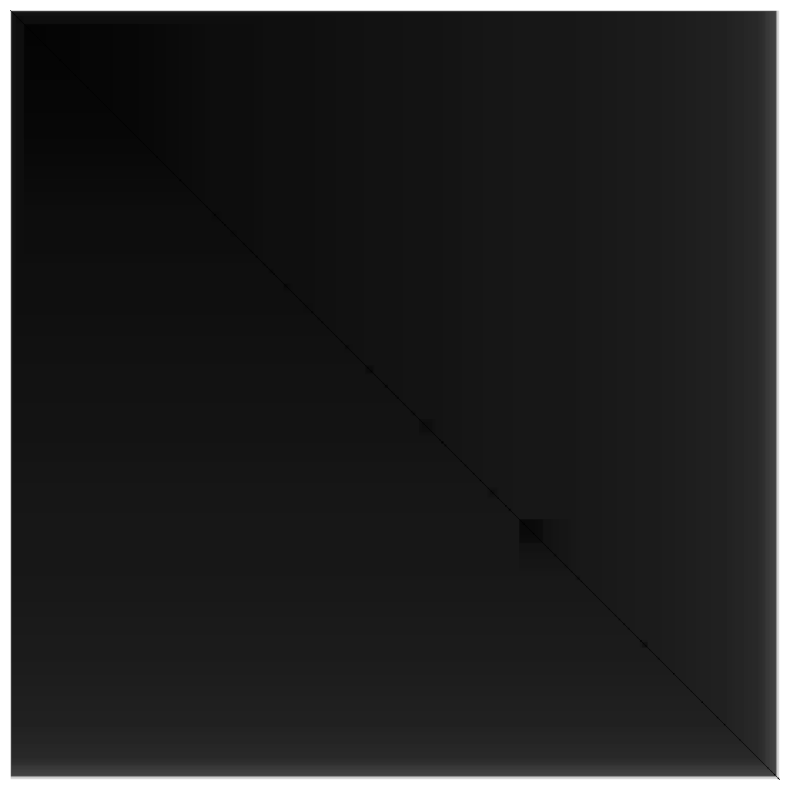}
        \caption{B-PCA-20}
        \label{fig:VAT20PCA}
    \end{subfigure}
    \hfill
    \begin{subfigure}[b]{0.16\textwidth}
        \centering
        \includegraphics[width=\textwidth]{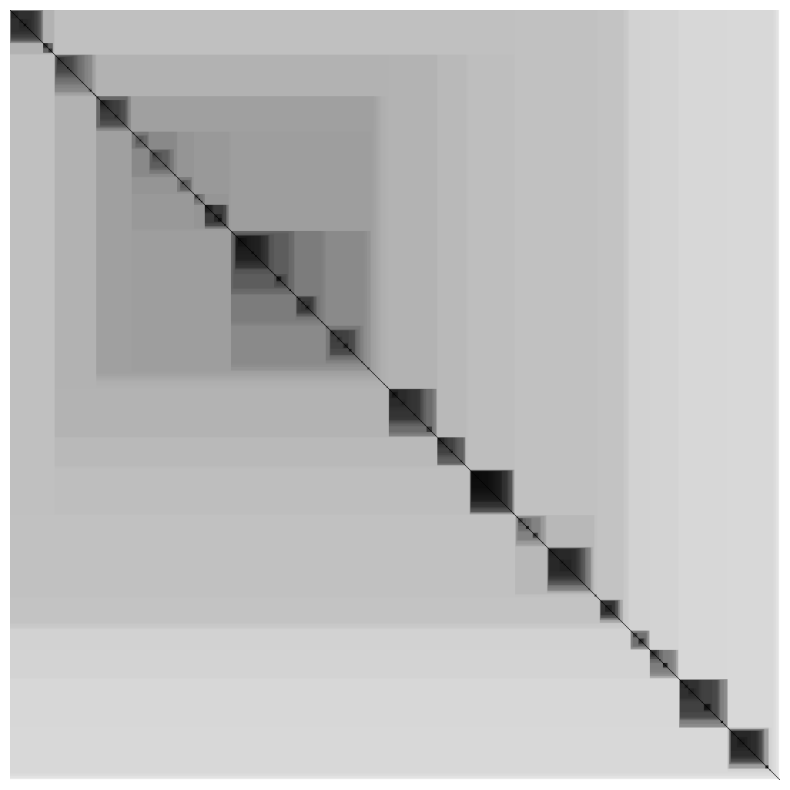}
        \caption{SSL-RN-20}
        \label{fig:VAT-RN}
    \end{subfigure}
    \hfill
    \begin{subfigure}[b]{0.16\textwidth}
        \centering
        \includegraphics[width=\textwidth]{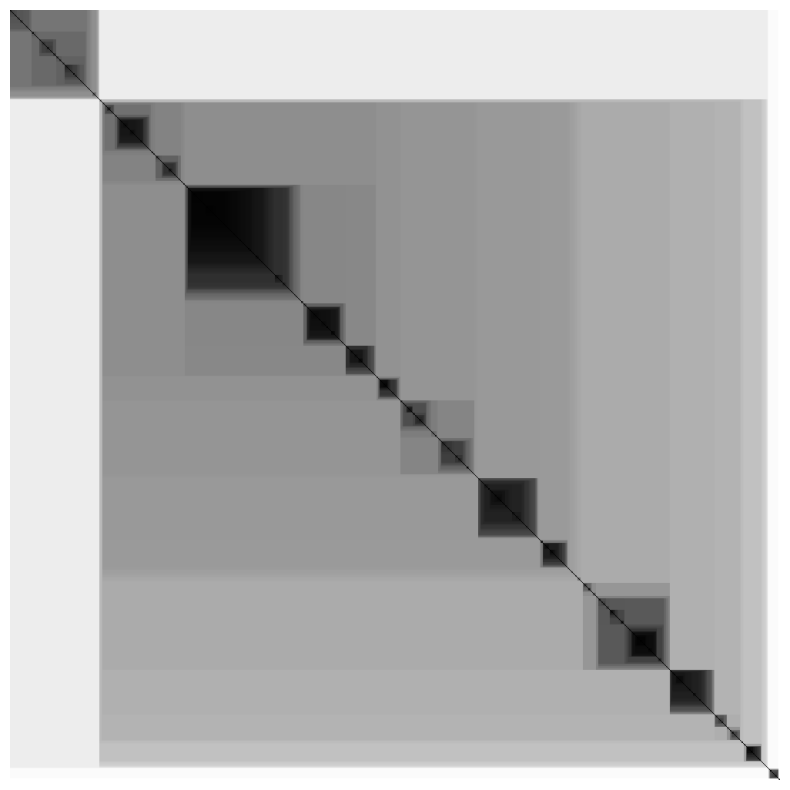}
        \caption{SSL-VGG-20}
        \label{fig:VAT-VGG}
    \end{subfigure}
    \hfill
    \begin{subfigure}[b]{0.16\textwidth}
        \centering
        \includegraphics[width=\textwidth]{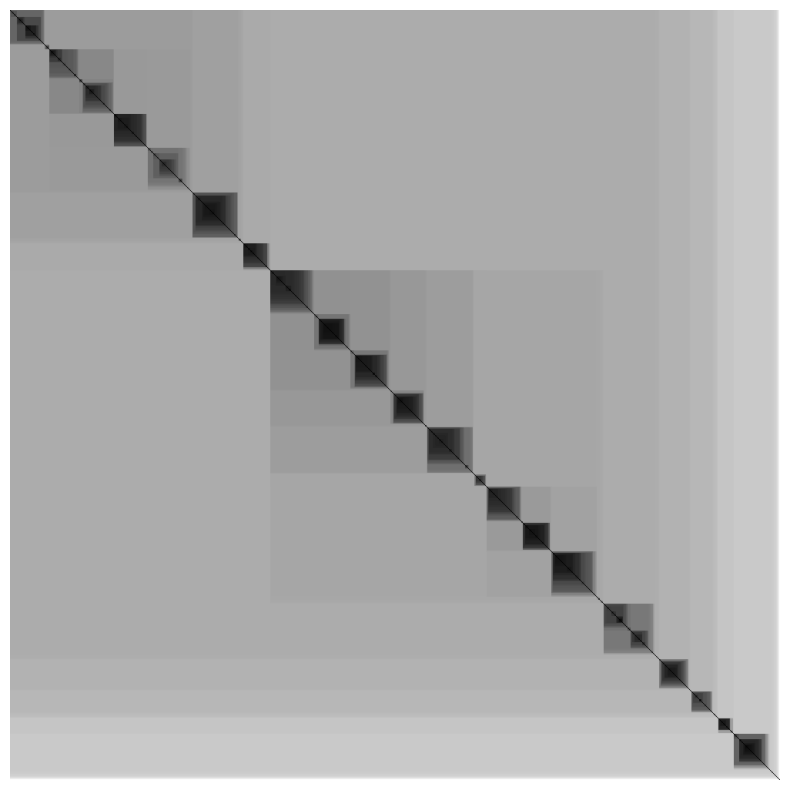}
        \caption{SSL-ViT-20}
        \label{fig:VAT-ViT}
    \end{subfigure}
    \hfill
    \begin{subfigure}[b]{0.16\textwidth}
        \centering
        \includegraphics[width=\textwidth]{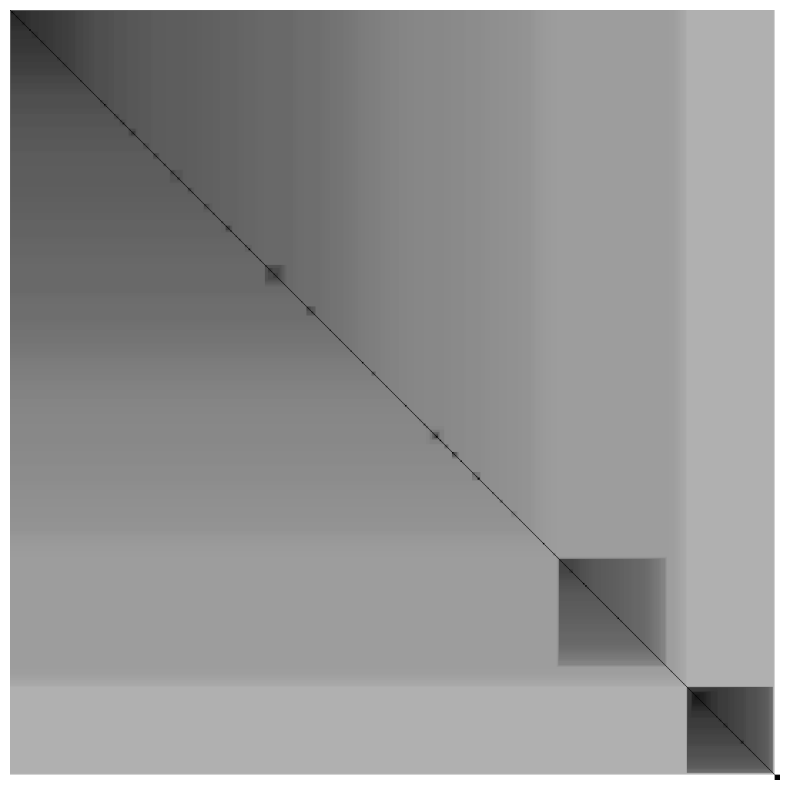}
        \caption{B-AE-20}
        \label{fig:VAT-AE}
    \end{subfigure}
    
    \caption{VAT plots of the different feature vectors.}
    \label{fig:VATs}
\end{figure*}

\begin{figure}[hbt!]

    \centering
    \includegraphics[width=\columnwidth]{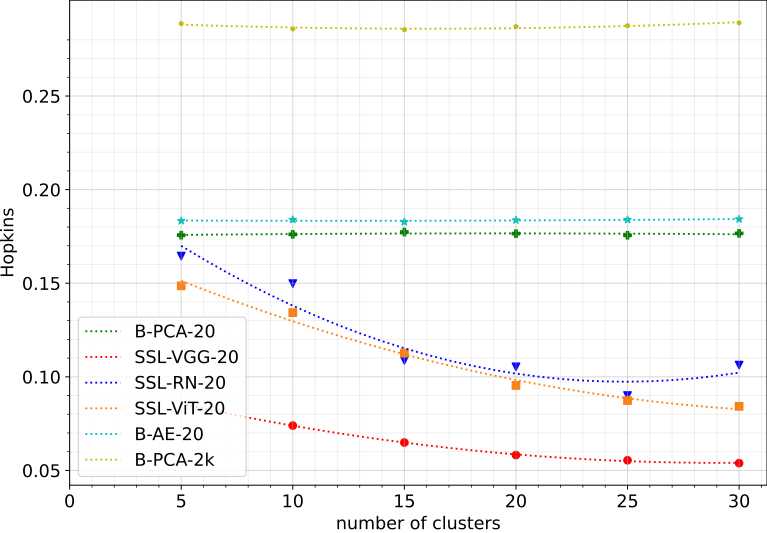}
        \caption{Hopkins scores of the extracted features.}
    \label{fig:Hopkins}
    
\end{figure}

\section{Experimental results} \label{sec:Results}

We started the performance evaluation with the analysis of the feature space of a few experimentally trained models to provide an insight about the content dynamics of the specific range of PCA components, before we continued with the investigation of the number of clusters in the dataset, and set the number of PCA components to be used in clustering. Using the methodology proposed in previous section, we carried out a number of experiments to compare the performance of SSL and baseline models, and eventually we showcased the suitability of SSL models for two representative use cases.

\subsection{Effect of the dimensionality reduction} \label{sec:ModelParameters}
For configuring the PCA-based dimensionality reduction of the proposed self-supervised and baseline models, we first explored the cumulative sum of the EVR of the features, as shown in Figure~\ref{fig:VarRatio}, for the different models. Figure~\ref{fig:PCAonCNN} plots EVR of {the three SSL models and the baseline AE model.}
It can be seen that $90\%$ of the EVR is contained in at least $17$ components across the {SSL} models and that there is no significant change in the EVR when considering more than $20$ components. Therefore we selected the first 20 PCA components as the size of the vectors encoding the input images, which are also input for the K-means algorithm in the unsupervised branch in Figure~\ref{fig:ArchitectureDiagram}. {The baseline AE model shows comparable performance by encoding the same EVR of 90\% in 23 PCA components. In our experimental investigations, we discovered an intriguing result regarding the AE model in conjunction with PCA dimensions. Specifically, we found that there is an almost negligible difference in the evaluation metrics when using either 20 or 25 PCA dimensions. To facilitate more convenient and equitable comparisons in our subsequent experiments, we have opted to employ a 20-dimensional PCA-transformed embedding space also for the AE model, denoted as B-AE-20, aligning it with the SSL models.}


The plot in Figure~\ref{fig:DirectPCAonImages} represents the feature space for the simple, PCA baseline model. It can be seen that direct PCA on images requires 1752 components to keep 90\% EVR independent of the final number of clusters. Taking this into consideration we used two baseline models, one with 1752 and one with 20 (i.e. the same as with SSL models) PCA components, and we refer to them in the following as B-PCA-2k and B-PCA-20, respectively.

Comparison of results in Figures~\ref{fig:PCAonCNN} and~\ref{fig:DirectPCAonImages} shows that SSL {and AE} models are able to learn to encode the relevant information for cluster development in {roughly} 1\% of the components required by the baseline PCA model for the same 90\% EVR. Thus, {the SSL-based and the AE-based feature extraction} provide significant simplification of further processing because of the dimensionality reduction by two orders of magnitude.

\subsection{Clustering tendency of the features} \label{sec:FeaturesEval}
Using {the features from the models} derived in the previous section, {which are the} inputs to the K-means {algorithm}, we perform quantitative evaluation of how suitable are the resulting features for clustering.

The Hopkins scores evaluating the quality of the features using the SSL versus the baseline models are presented in Figure \ref{fig:Hopkins} for 5, 10, 15, 20 , 25 and 30 clusters. We also plot fitted curves for each model to emphasize the general trends. For the baseline models B-PCA-20, B-PCA-2k {and B-AE-20}, the clusterability is independent of the number of clusters so we expect to have constant values. The small variations of the {corresponding values} are due to random initialization of the Hopkins calculation procedure. This is not the case for the {SSL} models since the number of clusters is always the same as the number of classes of the CNN output, thus affecting the {overall} architecture.


The SSL-VGG-20 model has consistently the best performance across the different number of clusters. The scores of the proposed SSL-RN-20 model are close to the B-PCA-20 model for small number of clusters, and they improve as the number of clusters becomes larger than 10 where they are close to the scores of the SSL-ViT-20 model.

For a more comprehensive evaluation, we opted to focus on SSL models configured with 23 clusters, a choice that aligns with the optimal range for SSL architectures. Notably, this cluster count variation had no discernible impact on the baseline models. Subsequently, we conducted an analysis of VAT plots for both the selected SSL and baseline models, as depicted in Figure~\ref{fig:VATs}.
Figure~\ref{fig:VATVarRatio95} corresponding to B-PCA-2k contains no grouping at all, meaning that the feature vectors provided by this model are randomly distributed in the feature space. Figure~\ref{fig:VAT20PCA} corresponding to B-PCA-20 shows that there is only a small grouping (outlayers probably) in the lower right corner, which justifies the better performance compared to the B-PCA-2k on the Hopkins evaluation. {Another observation is that almost the entire matrix is dark, meaning that all of the samples are closely correlated, which can be explained by the 20 dimensions of the PCA-transformed feature space containing less than 50\% of the EVR. The distances between the samples are thus much smaller compared to the SSL models.} 

\begin{figure}[hbt!]

\begin{subfigure}{\columnwidth}
\centering
\includegraphics[width=\columnwidth]{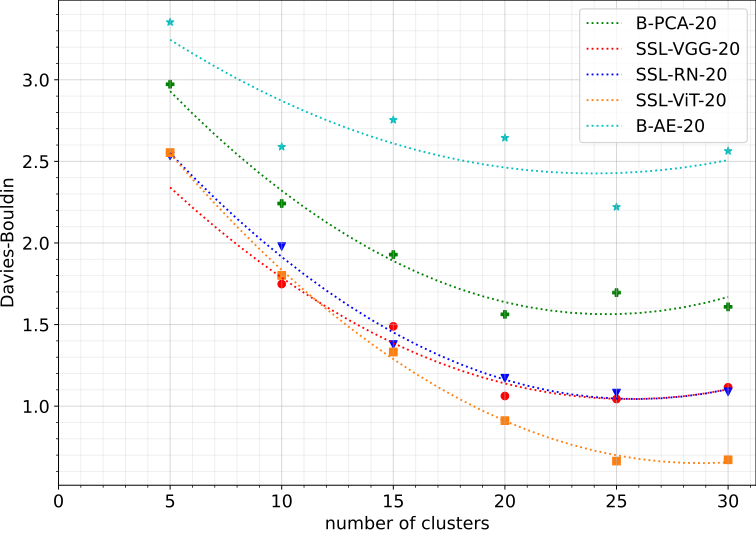}
\caption{{Davies-Bouldin scores}}
\label{fig:Davies-Bouldin}
\end{subfigure}
\begin{subfigure}{\columnwidth}
\centering
\includegraphics[width=\columnwidth]{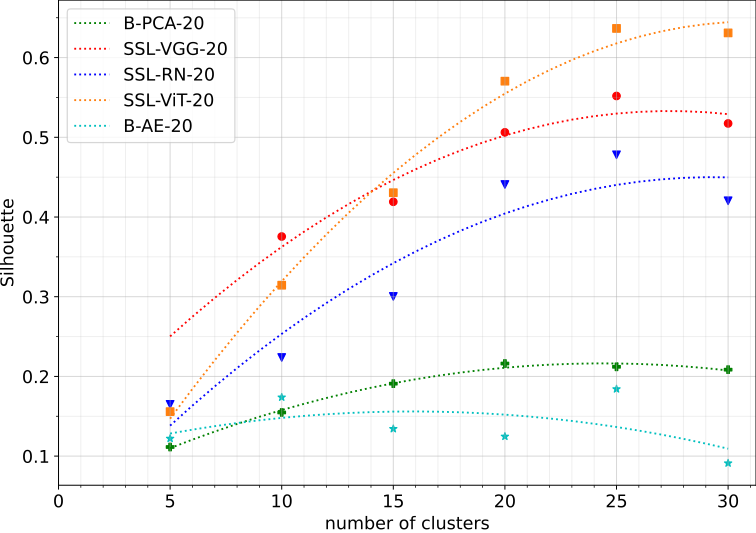}
\caption{{Silhouette scores}}
\label{fig:Silhouette}
\end{subfigure}

\caption{{Evaluation curves of metrics for quality of clustering.}}
\label{fig:ClusteringQuality}

\end{figure}

Figure~\ref{fig:VAT-RN} shows the clustering tendency of the features obtained with SSL-RN-20. There are around 20 distinguishable squares corresponding to groups of highly correlated samples. This confirms good scores on the Hopkins evaluation achieved by SSL-RN-20 models with 20 to 25 clusters. For the SSL-VGG-20 {in Figure~\ref{fig:VAT-VGG}}, the feature space contains three well separated, highly correlated groups of features and comparable size and number of darker squares as for the SSL-RN-20. The three well separated groups are the reason for the consistently better Hopkins scores of the SSL-VGG-20 compared to others, considering the Hopkins's definition. However, the number and size of visually distinguishable dark squares of the SSL-VGG-20 is comparable with the SSL-RN-20, suggesting that similar number of clusters could be existing in both feature spaces. {Regarding the SSL-ViT-20 in Figure~\ref{fig:VAT-ViT}, there are also approximately 20 clusters, but with very similar sizes. This suggests that the ViT may either  be able to focus on other specifics of the signals compared to CNNs or could be forcing the separation of the feature space by the defined number of clusters, regardless of the actual distribution.}

{Finally, B-AE-20 has clustering tendency scores somewhere between the SSL models and the simple PCA baseline models. There are three distinguishable clusters on the VAT plot, which is much lower than in the case of the SSL models. Considering the size of the squares, they seem to be more related to the general background features than the actual transmission patterns.}

From this section we can conclude that the PCA-only baseline models are not capable of encoding significant number of groups of correlated features from the input spectrograms. Following this conclusion, we only used the B-PCA-20 for further comparisons, since the B-PCA-2k feature vectors showed no clustering tendency at all, making their further evaluation irrelevant.

\subsection{Evaluation of clusters quality}
\label{sec:ClusteringEval}

Evaluating the clustering quality is performed according to the methodology described in Section~\ref{sec:EvalClus}. For the Davies-Bouldin scores plotted in Figure~\ref{fig:Davies-Bouldin} for 5, 10, 15, 20 , 25 and 30 clusters, it can be seen that the SSL models, i.e. SSL-RN-20, SSL-VGG-20 {and SSL-ViT-20} have consistently better scores compared to the baseline model B-PCA-20 {and B-AE-20}. Considering the SSL models only, the proposed SSL-RN-20 shows comparable {performance with the other two models, with SSL-ViT-20 being slightly better above 20 clusters.}

Here, again we plot fitted curves to show the trends for both metrics.
In general, the Silhouette score evaluation indicates the same pattern as the Davies-Bouldin score. The proposed SSL-RN-20 shows comparable performance {with the} SSL-VGG-20 {and SSL-ViT-20,} and significantly better than {the baselines}. This observation supports our conclusion from the previous section \ref{sec:FeaturesEval}, which evaluated the clustering tendency of the features, that {all three of the} SSL models have comparable groupings in the feature space and are significantly better than {both baseline models.}

\begin{table*}[hbt!]
	\centering
	\scriptsize
		\caption{{Statistics of the clusters. Legend: yellow-Stripes, red-Dotted, blue-Edges, gray-Idle, green-High intensity, white-Undefined}}
		\label{tab:ClustersMeta}
		\begin{tabular}{ccccccccccccccccccc}
			\toprule
			
			\multicolumn{3}{c}{B-PCA-20}
			& \phantom{}
			& \multicolumn{3}{c}{SSL-VGG-20}
			& \phantom{}
			& \multicolumn{3}{c}{SSL-RN-20}
                & \phantom{}
			& \multicolumn{3}{c}{SSL-ViT-20}
                & \phantom{}
			& \multicolumn{3}{c}{B-AE-20}
   
			\\\cmidrule{0-2}\cmidrule{5-7}\cmidrule{9-11}\cmidrule{13-15}\cmidrule{17-19}

                ID 
                & size \% 
                & Image mse.
                & 
                & ID 
                & size \% 
                & Image mse
                & 
                & ID 
                & size \% 
                & Image mse
                & 
                & ID 
                & size \% 
                & Image mse
                & 
                & ID 
                & size \% 
                & Image mse \\
                \midrule
                1 & 30.32 & 0.00395 && \cellcolor{gray!25}4 & \cellcolor{gray!25}14.0 & \cellcolor{gray!25}0.00007 && \cellcolor{yellow!25}9 & \cellcolor{yellow!25}6.28 & \cellcolor{yellow!25}0.00032 && \cellcolor{blue!25}11 & \cellcolor{blue!25}6.08 & \cellcolor{blue!25}0.00055 && \cellcolor{gray!25}1 & \cellcolor{gray!25}11.16 & \cellcolor{gray!25}0.00012 \\
                7 & 10.2 & 0.00267 && \cellcolor{yellow!25}5 & \cellcolor{yellow!25}7.23 & \cellcolor{yellow!25}0.00016 && \cellcolor{yellow!25}10 & \cellcolor{yellow!25}6.19 & \cellcolor{yellow!25}0.00034 && 1 & 5.78 & 0.00090 && \cellcolor{blue!25}3 & \cellcolor{blue!25}8.5 & \cellcolor{blue!25}0.00057 \\
                18 & 8.97 & 0.00420 && \cellcolor{blue!25}17 & \cellcolor{blue!25}5.66 & \cellcolor{blue!25}0.00017 && \cellcolor{gray!25}17 & \cellcolor{gray!25}6.1 & \cellcolor{gray!25}0.00006 && \cellcolor{gray!25}20 & \cellcolor{gray!25}5.24 & \cellcolor{gray!25}0.00029 && \cellcolor{blue!25}18 & \cellcolor{blue!25}7.98 & \cellcolor{blue!25}0.00021 \\
                9 & 3.75 & 0.00426 && \cellcolor{yellow!25}3 & \cellcolor{yellow!25}5.34 & \cellcolor{yellow!25}0.00016 && \cellcolor{red!25}8 & \cellcolor{red!25}5.77 & \cellcolor{red!25}0.00033 && \cellcolor{yellow!25}5 & \cellcolor{yellow!25}4.96 & \cellcolor{yellow!25}0.00070 && \cellcolor{gray!25}6 & \cellcolor{gray!25}7.07 & \cellcolor{gray!25}0.00011 \\
                6 & 3.66 & 0.00435 && \cellcolor{yellow!25}6 & \cellcolor{yellow!25}5.34 & \cellcolor{yellow!25}0.00015 && \cellcolor{yellow!25}1 & \cellcolor{yellow!25}5.75 & \cellcolor{yellow!25}0.00033 && \cellcolor{yellow!25}7 & \cellcolor{yellow!25}4.85 & \cellcolor{yellow!25}0.00059 && \cellcolor{yellow!25}19 & \cellcolor{yellow!25}6.39 & \cellcolor{yellow!25}0.00033  \\
                22 & 3.49 & 0.00447 && \cellcolor{yellow!25}15 & \cellcolor{yellow!25}5.12 & \cellcolor{yellow!25}0.00016 && \cellcolor{yellow!25}5 & \cellcolor{yellow!25}5.61 & \cellcolor{yellow!25}0.00036 && \cellcolor{yellow!25}0 & \cellcolor{yellow!25}4.81 & \cellcolor{yellow!25}0.00077 && \cellcolor{blue!25}4 & \cellcolor{blue!25}4.44 & \cellcolor{blue!25}0.00042 \\
                10 & 3.4 & 0.00432 && \cellcolor{yellow!25}12 & \cellcolor{yellow!25}5.08 & \cellcolor{yellow!25}0.00016 && \cellcolor{yellow!25}13 & \cellcolor{yellow!25}5.48 & \cellcolor{yellow!25}0.00033 && \cellcolor{yellow!25}15 & \cellcolor{yellow!25}4.7 & \cellcolor{yellow!25}0.00067 && \cellcolor{yellow!25}20 & \cellcolor{yellow!25}4.08 & \cellcolor{yellow!25}0.00049 \\
                16 & 3.19 & 0.00441 && \cellcolor{red!25}18 & \cellcolor{red!25}4.67 & \cellcolor{red!25}0.00035 && \cellcolor{blue!25}4 & \cellcolor{blue!25}5.48 & \cellcolor{blue!25}0.00010 && \cellcolor{yellow!25}13 & \cellcolor{yellow!25}4.67 & \cellcolor{yellow!25}0.00076 && \cellcolor{red!25}7 & \cellcolor{red!25}4.07 & \cellcolor{red!25}0.00082\\
                12 & 3.16 & 0.00452 && \cellcolor{red!25}1 & \cellcolor{red!25}4.57 & \cellcolor{red!25}0.00036 && \cellcolor{blue!25}2 & \cellcolor{blue!25}5.12 & \cellcolor{blue!25}0.00043 && \cellcolor{yellow!25}2 & \cellcolor{yellow!25}4.6 & \cellcolor{yellow!25}0.00072 && \cellcolor{yellow!25}9 & \cellcolor{yellow!25}4.02 & \cellcolor{yellow!25}0.00039 \\
                8 & 3.14 & 0.00433 && \cellcolor{blue!25}13 & \cellcolor{blue!25}4.52 & \cellcolor{blue!25}0.00055 && \cellcolor{yellow!25}6 & \cellcolor{yellow!25}5.1 & \cellcolor{yellow!25}0.00040 && \cellcolor{gray!25}10 & \cellcolor{gray!25}4.52 & \cellcolor{gray!25}0.00011 && \cellcolor{blue!25}16 & \cellcolor{blue!25}3.86 & \cellcolor{blue!25}0.00213 \\
                13 & 2.88 & 0.00429 && \cellcolor{yellow!25}8 & \cellcolor{yellow!25}4.44 & \cellcolor{yellow!25}0.00018 && \cellcolor{yellow!25}14 & \cellcolor{yellow!25}4.76 & \cellcolor{yellow!25}0.00037 && \cellcolor{yellow!25}18 & \cellcolor{yellow!25}4.5 & \cellcolor{yellow!25}0.00069 && \cellcolor{yellow!25}15 & \cellcolor{yellow!25}3.56 & \cellcolor{yellow!25}0.00038 \\
                14 & 2.84 & 0.00451 && \cellcolor{blue!25}9 & \cellcolor{blue!25}4.38 & \cellcolor{blue!25}0.00042 && \cellcolor{yellow!25}21 & \cellcolor{yellow!25}4.73 & \cellcolor{yellow!25}0.00047 && \cellcolor{yellow!25}8 & \cellcolor{yellow!25}4.46 & \cellcolor{yellow!25}0.00091 && \cellcolor{yellow!25}11 & \cellcolor{yellow!25}3.51 & \cellcolor{yellow!25}0.00066 \\
                19 & 2.76 & 0.00441 && \cellcolor{blue!25}0 & \cellcolor{blue!25}4.09 & \cellcolor{blue!25}0.00008 && \cellcolor{yellow!25}15 & \cellcolor{yellow!25}4.57 & \cellcolor{yellow!25}0.00029 && \cellcolor{yellow!25}19 & \cellcolor{yellow!25}4.45 & \cellcolor{yellow!25}0.00083 && \cellcolor{red!25}12 & \cellcolor{red!25}3.5 & \cellcolor{red!25}0.00062 \\
                17 & 2.7 & 0.00414 && \cellcolor{yellow!25}14 & \cellcolor{yellow!25}3.82 & \cellcolor{yellow!25}0.00017 && \cellcolor{yellow!25}19 & \cellcolor{yellow!25}4.5 & \cellcolor{yellow!25}0.00057 && \cellcolor{yellow!25}6 & \cellcolor{yellow!25}4.35 & \cellcolor{yellow!25}0.00075 && \cellcolor{yellow!25}21 & \cellcolor{yellow!25}3.48 & \cellcolor{yellow!25}0.00040 \\
                21 & 2.68 & 0.00463 && \cellcolor{yellow!25}19 & \cellcolor{yellow!25}3.75 & \cellcolor{yellow!25}0.00017 && \cellcolor{gray!25}16 & \cellcolor{gray!25}3.94 & \cellcolor{gray!25}0.00010 && \cellcolor{yellow!25}9 & \cellcolor{yellow!25}4.28 & \cellcolor{yellow!25}0.00108 && \cellcolor{red!25}5 & \cellcolor{red!25}3.36 & \cellcolor{red!25}0.00116 \\
                20 & 2.28 & 0.00461 && \cellcolor{blue!25}10 & \cellcolor{blue!25}3.3 & \cellcolor{blue!25}0.00040 && \cellcolor{yellow!25}3 & \cellcolor{yellow!25}3.7 & \cellcolor{yellow!25}0.00015 && \cellcolor{yellow!25}16 & \cellcolor{yellow!25}4.24 & \cellcolor{yellow!25}0.00067 && \cellcolor{yellow!25}17 & \cellcolor{yellow!25}3.31 & \cellcolor{yellow!25}0.00060 \\
                11 & 2.04 & 0.00461 && \cellcolor{red!25}16 & \cellcolor{red!25}3.29 & \cellcolor{red!25}0.00035 && \cellcolor{yellow!25}20 & \cellcolor{yellow!25}3.26 & \cellcolor{yellow!25}0.00036 && \cellcolor{yellow!25}3 & \cellcolor{yellow!25}4.18 & \cellcolor{yellow!25}0.00062 && \cellcolor{yellow!25}8 & \cellcolor{yellow!25}3.17 & \cellcolor{yellow!25}0.00070 \\
                15 & 2.04 & 0.00463 && \cellcolor{red!25}2 & \cellcolor{red!25}3.22 & \cellcolor{red!25}0.00039 && \cellcolor{yellow!25}22 & \cellcolor{yellow!25}3.11 & \cellcolor{yellow!25}0.00045 && \cellcolor{yellow!25}12 & \cellcolor{yellow!25}4.08 & \cellcolor{yellow!25}0.00076 && \cellcolor{green!25}14 & \cellcolor{green!25}2.89 & \cellcolor{green!25}0.00193 \\
                5 & 1.98 & 0.00448 && \cellcolor{blue!25}11 & \cellcolor{blue!25}3.02 & \cellcolor{blue!25}0.00108 && \cellcolor{yellow!25}11 & \cellcolor{yellow!25}2.34 & \cellcolor{yellow!25}0.00036 && \cellcolor{yellow!25}21 & \cellcolor{yellow!25}4.03 & \cellcolor{yellow!25}0.00072 && \cellcolor{yellow!25}13 & \cellcolor{yellow!25}2.78 & \cellcolor{yellow!25}0.00031 \\
                4 & 1.87 & 0.00465 && \cellcolor{green!25}22 & \cellcolor{green!25}2.24 & \cellcolor{green!25}0.00318 && \cellcolor{yellow!25}12 & \cellcolor{yellow!25}2.31 & \cellcolor{yellow!25}0.00041 && 17 & 4.01 & 0.00075 && \cellcolor{yellow!25}0 & \cellcolor{yellow!25}2.76 & \cellcolor{yellow!25}0.00031 \\
                0 & 1.64 & 0.00428 && \cellcolor{green!25}20 & \cellcolor{green!25}1.28 & \cellcolor{green!25}0.00309 && \cellcolor{green!25}18 & \cellcolor{green!25}2.11 & \cellcolor{green!25}0.00330 && \cellcolor{yellow!25}14 & \cellcolor{yellow!25}3.42 & \cellcolor{yellow!25}0.00065 && \cellcolor{yellow!25}22 & \cellcolor{yellow!25}2.6 & \cellcolor{yellow!25}0.00123 \\
                3 & 0.84 & 0.00472 && 21 & 0.88 & 0.00041 && \cellcolor{yellow!25}0 & \cellcolor{yellow!25}2.07 & \cellcolor{yellow!25}0.00030 && \cellcolor{blue!25}22 & \cellcolor{blue!25}3.09 & \cellcolor{blue!25}0.00016 && \cellcolor{yellow!25}10 & \cellcolor{yellow!25}2.55 & \cellcolor{yellow!25}0.00032 \\
                2 & 0.18 & 0.00281 && \cellcolor{red!25}7 & \cellcolor{red!25}0.76 & \cellcolor{red!25}0.00045 && \cellcolor{gray!25}7 & \cellcolor{gray!25}1.7 & \cellcolor{gray!25}0.00006 && \cellcolor{blue!25}4 & \cellcolor{blue!25}0.7 & \cellcolor{blue!25}0.00025 && \cellcolor{green!25}2 & \cellcolor{green!25}0.96 & \cellcolor{green!25}0.00215 \\
                \midrule
                
                \multicolumn{2}{l|}{Image mse mean} & 0.00427 &&&& 0.00046 &&&& \textbf{0.00044} &&&& 0.00065 &&&& 0.00071\\
                \multicolumn{2}{l|}{Image mse std} & 0.00051 &&&& 0.00074 &&&& 0.00062 &&&& \textbf{0.00023} &&&& 0.00059\\
			\bottomrule
   
		\end{tabular}	
\end{table*}

\subsection{Analysis of clusters}
According to Sections~\ref{sec:FeaturesEval} and~\ref{sec:ClusteringEval}, models with 20 to 25 clusters are providing the best results on the used dataset. Considering this advantage of using the SSL architectures over the baseline clustering approach, we present two use cases where the learning capabilities of such model are being exploited: 
\begin{enumerate}
    \item Transmission detection, where we group the activity-rich clusters containing transmissions as one class, and the clusters with spectrograms that do not contain transmissions as another class.
    \item High granularity clustering, considering all of the existing shapes appearing in the formed clusters.
\end{enumerate}

\begin{figure*}[hbt!]
    \begin{subfigure}{\textwidth}
        \centering
        \includegraphics[width=\columnwidth]{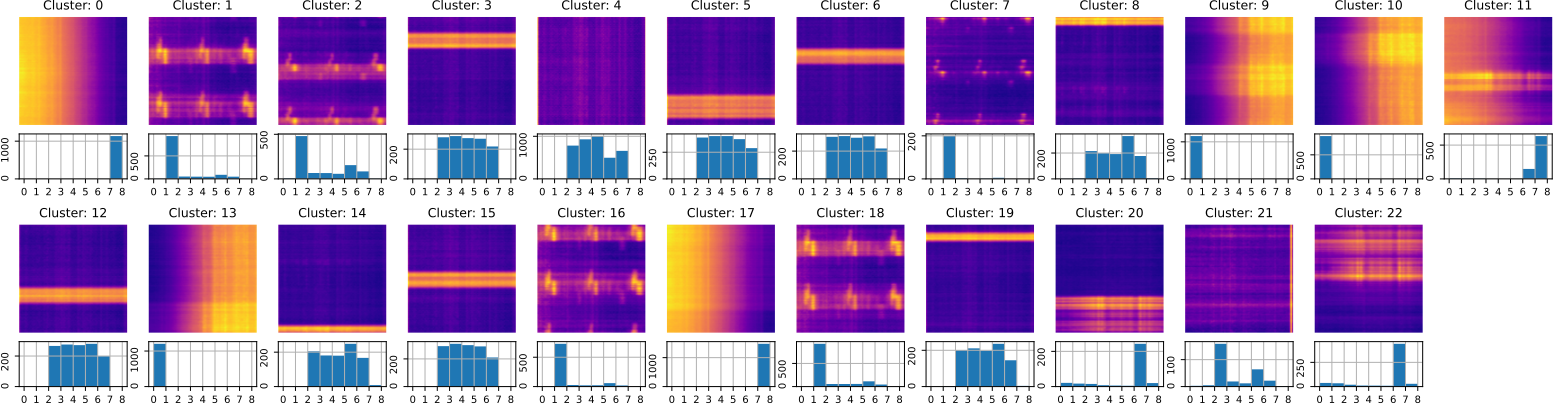}
        \caption{SSL-VGG-20}
        \label{fig:AvgHistVGG}
    \end{subfigure}
    \begin{subfigure}{\textwidth}
        \centering
        \includegraphics[width=\columnwidth]{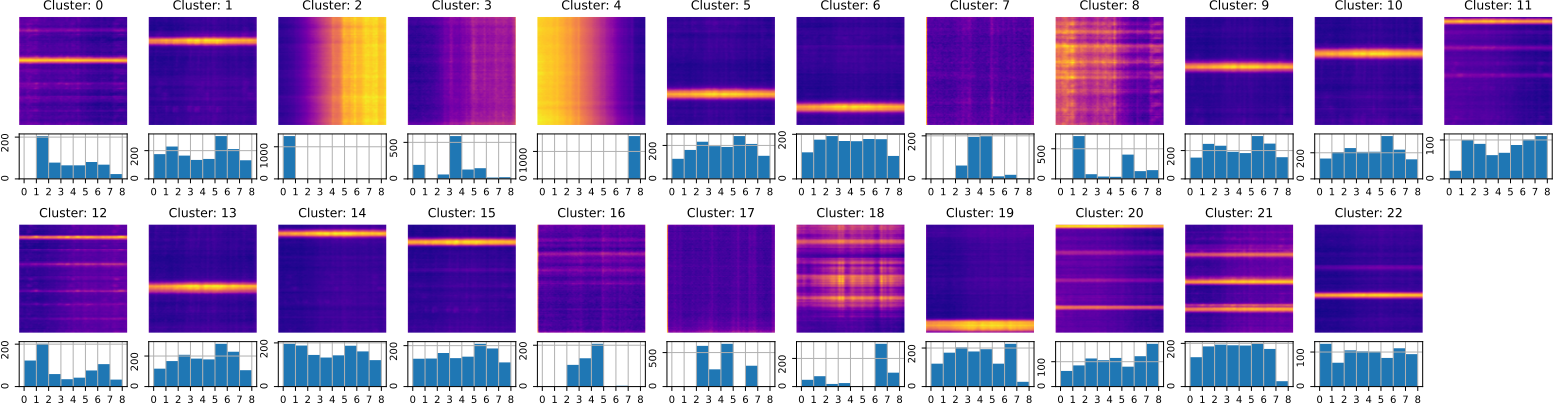}
        \caption{SSL-RN-20}
        \label{fig:AvgHistRN}
    \end{subfigure}
    \begin{subfigure}{\textwidth}
        \centering
        \includegraphics[width=\columnwidth]{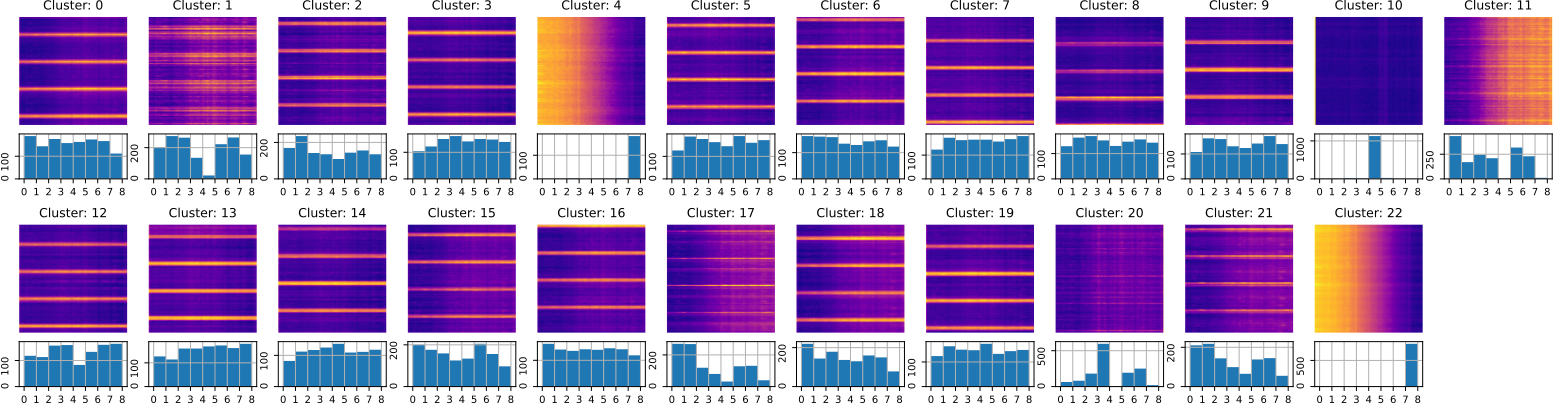}
        \caption{SSL-ViT-20}
        \label{fig:AvgHistVit}
    \end{subfigure}
    \begin{subfigure}{\textwidth}
        \centering
        \includegraphics[width=\columnwidth]{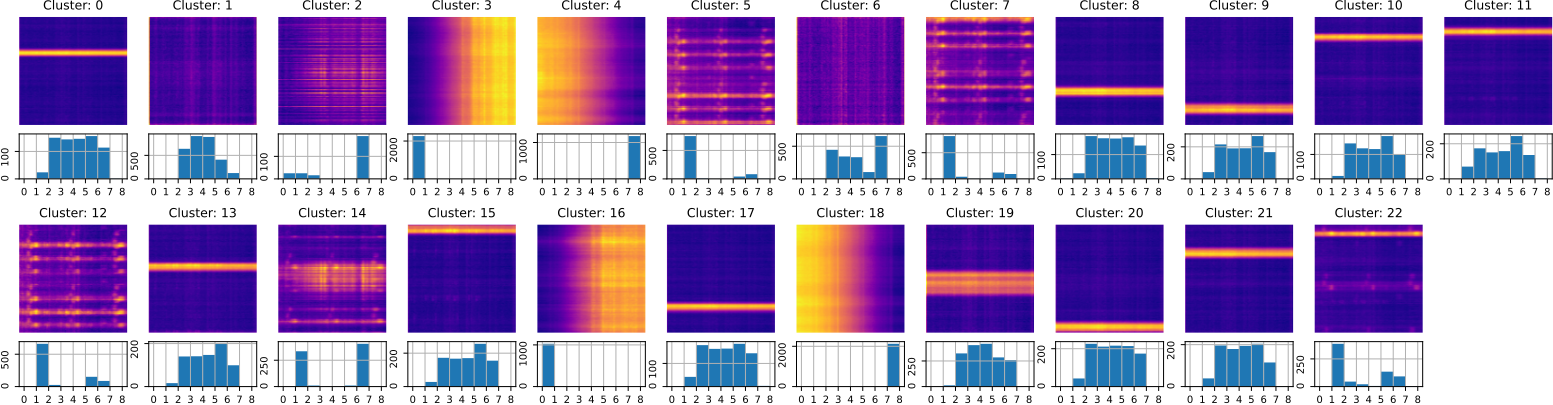}
        \caption{B-AE-20}
        \label{fig:AvgHistAE}
    \end{subfigure}
    \caption{Average spectrograms and frequency histograms of appearance for each cluster.}
    \label{fig:AvgAndHist}
\end{figure*}

Statistics of the three compared models are summarized in Table~\ref{tab:ClustersMeta}. Complementary to the table, Figure~\ref{fig:AvgAndHist} provides the average spectrograms and histograms of frequency sub-bands for the samples of each cluster. Cluster statistics in Table~\ref{tab:ClustersMeta} contain the cluster sizes in percentages of the entire dataset (\textit{Cluster size \%}) and image mean-squared-error similarities between each cluster's average spectrogram and the spectrograms of its assigned samples (\textit{Image mse}). {Colors of the cells are related to the clusters' content discussed later in Section \ref{sec:DiscusManual} and depicted in  Figure \ref{fig:AvgAndHist}.}

Interestingly, the B-PCA-20 has the lowest \textit{overall mean of inter-cluster (IC) distances}, one order of magnitude lower than the SSL models. But, this is a result of having small variance ratio contained in the 20 dimensions of the PCA transformed space of the raw images (baseline PCA approach), and to note again, using bigger amount of components resulted to even worse results in Section~\ref{sec:FeaturesEval}. On the other hand, the IC image distances are one order of magnitude worse compared to the SSL models, which confirms the performance seen in Section~\ref{sec:ClusteringEval}. 

SSL-VGG-20 has slightly better \textit{overall mean IC distance} than SSL-RN-20, which corresponds to better distinguishing of the features according to Figure~\ref{fig:Hopkins}. 

{Looking at the CNN-based models, SSL-VGG-20 and SSL-RN-20, they show almost identical performance for the \textit{Image mse mean}, 0.00046 and 0.00044, accordingly. At the same time, the overall \textit{Image mse std} is better for the SSL-RN-20.
While the SSL-ViT has the lowest \textit{Image mse std}, it has the worst \textit{Image mse mean} among the SSL models.} This means that although the encoded features are better grouped by the SSL-VGG-20 {and SSL-ViT-20, according to the metrics in Section \ref{sec:ClusteringEval}}, the differences between the clustered spectrograms {of these two SSL models} are actually larger than the ones resulting from the SSL-RN-20 model, thus showing that the latter has slight advantage in grouping wireless transmissions with similar shapes.
{The simple baseline model B-PCA-20, expectedly, has the worst performance, with \textit{Image mse mean} of 0.00427 which is one order of magnitude larger than the rest of the models.}

{Interestingly, the B-AE-20 has very close performance to the SSL models with \textit{Image mse mean} of 0.00071, contrary to the significant underperforming seen on the feature space evaluations with the other metrics in Figures~\ref{fig:Hopkins}, \ref{fig:Davies-Bouldin} and \ref{fig:Silhouette}. This means that, while the feature space is not very clearly separated, the clusters are formed by similar types of transmissions. However, it should be noted that the number of clusters for the experiment equal to 23 was selected based on the performance of the SSL models seen using the previous metrics. The evaluations of the B-AE-20 in the embedded domain show similar results almost for the entire range of the number of clusters. So even if the results are good in this case, selection of the right number of clusters for this model is a guesswork, and not a result of methodical reasoning.}

\subsection{Manual evaluation} 
\label{sec:DiscusManual}

Manual evaluation for the selected models with 23 clusters relies on the distribution of samples per cluster with the relevant cluster statistics in Table~\ref{tab:ClustersMeta} and the average spectrogram corresponding to each cluster in Figure~\ref{fig:AvgAndHist}. Averaging of spectrogram samples leads to highlighting the position of the most common types of activities that appear in each of the clusters. This visualization reveals which type of spectrum activity (i.e. transmissions of existing wireless technologies) corresponds to each of the clusters. So, instead of labeling large amounts of data, we only need to analyse the content of the formed clusters.
Since the B-PCA-20 model has shown weak clustering performance according to all previous metrics, manual evaluation was performed only on the clusters obtained with the SSL models {and the B-AE-20.}

Based on the activity shapes and frequency histograms, clusters {of all the} models can be separated in six major groups (with color marking as in Table~\ref{tab:ClustersMeta}): \textit{Idle} clusters (grey), \textit{Edge} clusters (blue), \textit{High intensity} activity (green), \textit{Dotted} activity (red), horizontal \textit{Stripes} activity (yellow), and \textit{Unidentified} (white).

{\textit{Idle} group} is represented by cluster 4 for SSL-VGG-20 and clusters 7, 16 and 17 for the SSL-RN-20, containing 14\% and 11.74\% of the data, respectively. This means that there is a slight difference in the number of samples assigned to no activity clusters between the models, {as discussed later in Section \ref{sec:LabeledEval}. The SSL-ViT-20 has two clusters, 20 and 10 with \textit{Idle} samples, totaling 9.76\% of the data, which is even less than the other two SSL models. Considering the averaged clusters in Figure~\ref{fig:AvgHistVit} we conclude that it is even more sensitive to the low power transmissions, so in Idle clusters are only the samples with containing very low or no transmissions. The B-AE-20 shows completely opposite behaviour, by grouping more than 18\% of the data as Idle, contained in the clusters 1 and 6. This means that it is the least sensitive model towards detection of low power transmissions. From practical point of view, the SSL-ViT-20 model would be desirable if there is high criteria for avoiding interference when estimating opportunities for transmission.}

The left \textit{Edges} and right \textit{Edges} clusters are mainly formed by samples from the edges (left-most and right-most) of the observed bandwidth (also shown on the histograms under each averaged spectrograms). The varying sensitivity of the sensor towards the edges of the bandwidth appears as highly influential information for {all of the} models. 

The SSL-VGG-20 has clusters 9, 10 and 13 from the left \textit{Edge} of the bandwidth, totaling 12.2\% of the whole data which matches nearly perfect with 1/8=12.5 of the data falling in the left-most sub-band (the entire bandwidth being separated in 8 sub-bands). The right \textit{Edge} cluster formed by clusters 0, 11 and 17, totaling 12.77\% of the data, follows the same pattern. Although it is evident that these clusters also contain additional content (e.g. clusters 9, 10, 13), it is clear that the most prominent shape is the varying intensity background noise. This means that for these bandwidth regions, SSL-VGG-20 is focused more on the background noise than the transmission activity.
For the SSL-RN-20, the left \textit{edge} and right \textit{Edge} spectrum samples are contained in only two clusters, labeled 2 and 4, with roughly equal size of 5.12\% and 5.61\% of the total data, respectively. This means that the varying noise is also important for the SSL-RN-20, but more weight is given to the wireless transmission activities.

{The SSL-ViT-20 has clean clusters for the right \textit{Edge}, 4 and 22 (totaling 3.79\% of the data), while the left \textit{Edge} cluster, 11, is mixed with other types of activities. Meaning, the edge characteristics are not decisive in the clustering process of this model.
The baseline B-AE-20 shows similar performance like the SSL-VGG-20 regarding the \textit{Edge} samples, with 24.78\% of the data falling into the clusters 3, 4, 16 and 18. Again, almost equal to the total amount of the samples from the edges.}

Considering the \textit{High intensity} activity samples, for SSL-VGG-20 they are distributed in clusters 20 and 22 (totaling 3.52\%) while for SSL-RN-20 they are only in cluster 18 (2.11\%). For this activity SSL-VGG-20 achieves better clustering. Based on the histogram of cluster 19 which has higher number of samples in the 7th sub-band, exactly where these activities appear according to the histogram of cluster 18, SSL-RN-20 is confusing the high intensity activity with the horizontal \textit{Stripes} activity in cluster 19. {The SSL-ViT-20 has no cluster with \textit{High intensity} activities, and the B-AE-20 has clusters 14 and 2 (containing 3.85\%), again very close to the numbers of the SSL-VGG-20 model.}

Looking at the \textit{Dotted} activity samples, contained in clusters 1, 2, 7, 16 and 18 for SSL-VGG-20 and in cluster 8 for SSL-RN-20, we observe that SSL-VGG-20 focuses much more on these patterns, ignoring the coexistence of other types of activities in the spectrograms. {The same pattern appears also for the B-AE-20 model, considering clusters 5, 7 and 12, which are slightly more mixed with other types of transmissions from the sub-bands 6 and 7.} The histograms show that samples from these clusters contain all of the samples of the second sub-band, including those belonging to the Horizontal stripes activity which should obviously appear across the entire bandwidth as indicated in Figure~\ref{fig:SampleData}. This is also confirmed by the histograms of the horizontal \textit{Stripes} activity, {for both SSL-VGG-20 and B-AE-20,} which are expected to be distributed along the entire bandwidth, but they are only in the sub-bands 3-7, confirming misclassifications of the \textit{Dotted} activity samples. 

On the other hand, SSL-RN-20 focuses much more on the horizontal \textit{Stripes} activity samples, forming many clusters based only on the vertical location of the activity and having well distributed histograms across all sub-bands, as one would expect. Most of the samples that contain the horizontal \textit{Stripes} and the \textit{Dotted} activity samples are thus classified in the clusters with the horizontal \textit{Stripes} (e.g. clusters 0, 11, 12 and 20). {For the SSL-ViT-20, most of the samples (70.6\%) fall into the horizontal \textit{Stripes} class, meaning this type of activity is the main focus for this model while only some of the edges and \textit{Idle} samples form separate clusters.}

{In general, SSL-VGG-20 and SSL-RN-20 show better results on the manual evaluation compared to the SSL-ViT-20. Surprisingly, the B-AE-20 results look even better than the SSL-ViT-20, because of having more variety in the clusters with different types of transmissions, and slightly worse than the SSL-VGG-20.} 

Both {SSL-VGG-20 and SSL-RN-20} architectures show comparable performance, {which is more balanced towards different types of transmission activities when compared to the SSL-ViT-20 and the B-AE-20}, having similar pros and cons. The difference is that they focus on different shapes of activities, so samples containing multiple transmissions are distributed across different clusters, predominantly the Dotted activity for SSL-VGG-20 and the Horizontal stripes activity for SSL-RN-20. {Considering this finding only these two models will be considered in the following use cases.}

\subsubsection{Use case 1 - Transmissions detection}
This use case is equivalent to binary classification of the labeled (positive) samples between the two groups of clusters which represent the Idle state and the Occupied state.

\paragraph{Idle state - No/weak spectrum activity}
According to Figure~\ref{fig:AvgAndHist}, clusters with labels 7, 16, 17 for SSL-RN-20 and 4 for SSL-VGG-20, contain no or very low activity (\textit{Idle}). These clusters represent the parts of spectrum where no transmissions are recorded, or the sources of the signals are very distant so the signal intensity is very low. In the context of spectrum availability, they could be considered very similar since in both cases there are no existing patterns of transmissions with significant intensity.

\paragraph{Occupied - presence of transmissions}
All other clusters, 20 for the SSL-RN-20 and 22 for the SSL-VGG-20, contain some type of activity, so in the sense of spectrum availability they represent occupied spectrum segments. By considering the counts in Table \ref{tab:ClustersMeta}, it can be seen that these amount to 88\% (SSL-RN-20) and 86\% (SSL-VGG-20) of spectrograms that contain at least one transmission. This represents valuable information about the spectrum occupancy that was extracted in automatic way and all the manual work is reduced to inspection of the averaged clusters' spectrograms.

\subsubsection{Use case 2 - Distinguishing types of activities} 
\label{sec:UseCase2}

The cluster contents and the cluster size allow for better understanding of types of activities and frequency of their appearance across the monitored radio spectrum. Considering only the low-activity clusters, we acknowledge the percentage of available slots for transmission during time. Analog to this, there are also percentages for other types of activities. This means that information is provided about what portion of the spectrum is occupied with certain types of activity or combination of overlapping activities.
This is useful when more granular separation of the spectrograms is required. The mentioned manual evaluation of the averaged spectrograms has a crucial role in such case. For this use case we only exemplify the SSL-RN-20 model with 23 clusters. 

The clusters containing different types of combined activities can be considered as separate classes. In this way, besides the information for the spectrum occupancy, additional information can be derived for the frequency of occurrence and types of overlapping transmissions. In summary, based on manual evaluation of the clusters in Section~\ref{sec:DiscusManual}, the number of activity types can be reduced to 6, including the \textit{Idle} type of clusters containing 11.74\% of the total data.

\paragraph{Occupied - Horizontal stripes}
According to the average spectrograms visualization in Figure~\ref{fig:AvgHistRN}, there are several clusters with mostly horizontal stripes (clusters 1, 5, 6, 9, 10, 13, 14, 15, 19 and 22), which account for 51.35\% of the dataset. These samples represent the same type of activity, namely IEEE 802.15.4 transmissions, that occupy the entire bandwidth covered with a single segment, and they last for short time with regards to the time duration of a single sample. The location of the stripe is varying in vertical direction because the segmentation of the data in non-overlapping windows is fixed, as explained previously in Section \ref{sec:Data}, without correlation to the appearance times of any of the activities.

\paragraph{Occupied state - Horizontal stripes + other types of activity}
Second type of activities is observed when the same horizontal stripes appear combined with other activities (clusters 0, 11, 12, 20, and 21), representing 14.71\% of the samples. Together they represent separate cluster of segments where multiple activities are occurring in the same spectrogram segment. Some of the activities we distinguish correspond to concurrent IEEE 802.15.4, proprietary and LoRA transmissions.

\paragraph{Occupied state - High intensity activity}
Cluster 18 (2.11\% of the data) contains \textit{High intensity} activity spread along the time axis (vertical) direction. These activities correspond to LoRA transmissions. Many of these transmissions are also coexisting with the IEEE 802.15.4 transmissions, as discussed in Section~\ref{sec:DiscusManual}.

\paragraph{Occupied state - Dotted activity}
Cluster 8 (5.77\% of the data) contains shapes of dotted activity that are apparent at various locations along the time (vertical) axis. Such activities correspond to LoRa and proprietary transmissions.

\paragraph{Edge sub-bands}
It is important to notice that varying background noise is also influential in the clustering. Samples in clusters 2 and 4 contain significant amount of background noise, having weaker values on one side of the spectrogram and gradually increasing in horizontal direction (frequency axis). As a consequence of this, such spectrograms are clustered separately making up 10.6\% of the data, which cannot be specified if it is occupied or not.
\begin{table*}[hbt!]
	\centering
	\footnotesize
		\caption{Transmission detection performance and complexity comparison.}
		\label{tab:Performance}
		\begin{tabular}{c|cccccc}
		\toprule
            Algorithm & \cellcolor{gray!25}SSL-RN-20 & SSL-VGG-20 & Dilate/erode \cite{gale2020automatic} & TX grouping \cite{gale2020automatic} & SSL-ViT-20 & B-AE-20\\
            \midrule
            Precision \% & \cellcolor{gray!25}76.0 & 76.6 & \textbf{77.7} & 68.2 & 74.1 & 76.8 \\
            Recall \% & \cellcolor{gray!25}93.6 & 94.1 & 86.3 & 94.7 & \textbf{96.5} & 90.3 \\
            F1 \% & \cellcolor{gray!25}{83.9} & \textbf{84.5} & 81.8 & 79.3 & 83.9 & 83.0 \\
            \midrule
            Num. of parameters & \cellcolor{gray!25}\textbf{11.2 M} & 114.1 M & / & / & 54.6 M & 19.9 M \\
            \bottomrule
   		\end{tabular}	
\end{table*}

\subsection{Evaluation with labeled data}
\label{sec:LabeledEval}
At last, we evaluate the performance of the models for transmissions detection according to Section~\ref{sec:EvalLabels}, feeding labeled samples of transmissions to the models and determining how many of these samples were correctly distributed to clusters with present transmissions and clusters with no transmissions. Table~\ref{tab:Performance} summarizes the evaluation. 
We also compare the proposed SSL-RN-20 solution with existing referenced transmission detection algorithms evaluated on the same dataset in \cite{gale2020automatic}. Surprisingly, the \textit{Dilate/Erode} achieves the highest precision score with 1.1pp (percentage point) margin and the lowest recall score with 6-7pp margin, while the \textit{TX grouping} achieves the second highest recall score and the lowest precision with almost 8pp margin. This means that, although the \textit{Dilate/Erode} has the highest ratio of real transmissions in the detected transmissions, it fails to detect significant amount (roughly 14\%) of the total number of transmissions. On the opposite, the \textit{TX grouping} detects the high percentage of the total number of transmissions, but at the same time also classifies many of the \textit{Idle} samples as samples with transmissions, causing poor precision.

Contrary to the \textit{Dilate/Erode} and \textit{TX grouping}, the SSL-based models show more balanced performance, considering the difference between the precision and recall, thus achieving higher F1 scores by margin of {more than} 2pp. In general, the SSL-based models outperform the referenced ones on the task of transmission detection in continuously sensed spectrum data.

{The B-AE-20, employed as a baseline model also outperforms the referenced models on this task, however it is still slightly worse than the SSL-based models.}

Now, comparing the SSL-based models, {all three} of them show comparable performance. {The (original) SSL-VGG-20 has slight advantage of up to 0.6pp in F1 score. And the SSL-ViT-20 has more significant 2.9pp advantage on the Recall. Thus, while CNN-based SSL models are good for fine grained, transmission-specific clustering they may confuse certain transmissions with background noise. On the other hand, the ViT-based model is able to better separate Idle samples, as can also be seen from Figure~\ref{fig:AvgAndHist}, thus achieving better Recall performance on the labelled evaluation for activity detection according to Table~\ref{tab:Performance}.}
Anyway, the proposed SSL-RN-20 achieves such performance by utilizing roughly 10 times less trainable parameters compared to the SSL-VGG-20, {roughly 5 times less compared to the SSL-ViT-20,} which could be a beneficial trade-of regarding the computing requirements for potential applications. {The scripts used for these experiments are publicly available as open source software\footnote{\url{https://github.com/sensorlab/self-supervised-spectrum-sensing}}}.

\section{Conclusions}
\label{sec:concl}
In the work presented in this paper, we investigated the suitability of SSL for automatic feature learning and clustering of real-world radio spectrum events making use of {three different} SSL {models, based on a similar architecture.} SSL-VGG-20 originally developed for RGB image feature learning, its adaptation to the specifics of wireless spectrograms SSL-RN-20,{ and SSL-ViT-20, an adaptation based on a vision transformer.} We employed these architectures to develop SSL models with no labels and no prior knowledge, and compared their performance against {two} baseline models, PCA dimensionality reduction {as a simple, yet informative representation learning approach and another, CNN-based auto-encoder, both complemented with} K-means clustering.
The comparison was first made for the evaluation of the clustering tendency of the input data with appropriately proposed metrics for the different models and the quality of the clustering results. The self-supervised models enable compact feature extraction, encoded in significantly lower number of dimensions (reduction is of two orders of magnitude) while preserving the same amount of EVR of the input features. They also outperform the baseline models in the clustering quality by 0.3 according to Silhouette scores, and by 0.6 according to Davies-Bouldin score. The study was supported also by manual evaluation based on visual inspection and evaluation by labeled data. Results show that SSL-based models outperform the existing transmission detection algorithms on continuously sensed data by more than 2pp according to F1 scores.

\section*{Acknowledgments}
This work was funded in part by the Slovenian Research Agency under the grant P2-0016.
This project has received funding from the European Union’s Horizon Europe Framework Programme under grant agreement No 101096456 (NANCY).
The project is supported by the Smart Networks and Services Joint Undertaking and its members.

\bibliographystyle{IEEEtran}
\bibliography{arxiv}

\end{document}